\documentclass[aps,12pt]{revtex4}

\usepackage{graphicx}
\usepackage{graphics}
\usepackage{color}
\usepackage[intlimits,tbtags]{amsmath}
\usepackage{amsfonts,amssymb,amsthm}

\def\beq{\begin{equation}}
\def\eeq{\end{equation}}
\def\bea{\begin{eqnarray}}
\def\eea{\end{eqnarray}}
\def\nn{\nonumber}

\begin{document}

\title{MPA for TASEP with a generalized update on a ring}

\author{B. L. Aneva~$^{\dag^1}$ and J. G. Brankov~$^{\dag^2\dag^3}$ \email{brankov@theor.jinr.ru}}

\affiliation{$^{\dag^1}$~Institute for Nuclear Research and Nuclear Energy, Bulgarian
Academy of Sciences, 1784 Sofia, Bulgaria}

\affiliation{$^{\dag^2}$~Bogoliubov Laboratory of Theoretical Physics, Joint
Institute for Nuclear Research, 141980 Dubna, Russia}

\affiliation{$^{\dag^3}$~Institute of Mechanics, Bulgarian
Academy of Sciences, 1113 Sofia, Bulgaria}

\begin{abstract}

We apply the Matrix Product Ansatz to study the Totally Asymmetric Simple Exclusion Process on a ring with a generalized discrete-time dynamics depending on two hopping probabilities, $p$ and $\tilde{p}$. The model contains as special cases the TASEP with parallel update, when $\tilde{p} =0$, and with sequential backward-ordered update, when $\tilde{p} =p$.

We construct a two-dimensional matrix-product representation  and use it to obtain exact finite-size expressions for the partition function, the current of particles and the two-point correlation function. Our main new result is the derivation of the finite-size pair correlation function. Its behavior is analyzed in different regimes of effective attraction and repulsion between the particles, depending on whether $\tilde{p} >p$ or $\tilde{p} < p$. In particular, we explicitly obtain an analytic expression for the pair correlation function in the limit of irreversible aggregation $\tilde{p}\rightarrow 1$, when the stationary configurations contain just one cluster.


\end{abstract}

\maketitle

\section{Introduction}

The asymmetric simple exclusion process (ASEP) is one of the simplest exactly solved models of driven many-particle systems with particle conserving bulk stochastic dynamics \cite{D98,S01}. In the extremely asymmetric case, when particles are allowed to move in one direction only, it reduces to the totally asymmetric simple exclusion process (TASEP). For its description in the context of interacting Markov processes we refer to \cite{S70}. In the course of time, ASEP and TASEP became paradigmatic models for understanding the broad variety of nonequilibrium phenomena. Devised to model kinetics of protein synthesis \cite{MGP68}, TASEP and its numerous extensions have found many applications to vehicular traffic flow \cite{NS92,CSS00,H01,Schad01}, biological transport \cite{PFF03,KVHD03,NKL04,LPVHDH,NKP13,BPB14,TKM15,CTK15},
one-dimensional surface growth \cite{KS88,S05}, forced motion of colloids in narrow channels \cite{CL99,K07}, spintronics \cite{RFF07}, transport of "data packets" on the Internet \cite{HBKSS01}, current through chains of quantum dots \cite{KO10}, limit order market \cite{WSC02}, to mention some.

The stationary properties of the original TASEP in continuous time were exactly obtained by different methods. The first exact solution was based on a recurrence relation, obtained at special values of the parameters in \cite{DDM92}, and generalized by Sch\"{u}tz and Domany \cite{SD93}. Using this recursion, closed expressions for the average occupations of all sites were obtained. The stationary states of TASEP and its generalization to a reaction-diffusion process with
two-site interactions, were studied also by using the quantum Hamiltonian formalism \cite{S94}. A combinatorial approach and mapping on weighted lattice paths shed new light on the formulation and solution of particle hopping models in continuous and discrete time \cite{DS05,BE12}.
An effective way to exploit the recursive properties of the steady states of a variety of one-dimensional processes is the matrix product ansatz (MPA). According to the MPA, the stationary configuration probabilities can be expressed as matrix elements of products of operators representing particles and holes. Such a matrix-product representation of the steady-state probability distribution for TASEP was found by Derrida, Evans, Hakim, and Pasquier \cite{DEHP}. Their formalism involves two square matrices, D and E, which are infinite-dimensional in the general case and satisfy a quadratic algebra, known as the DEHP algebra. Krebs and Sandow \cite{KS97} proved that the stationary state of any
one-dimensional system with random-sequential dynamics involving nearest-neighbor hopping and single-site boundary terms can always be written in a matrix-product form. Fock representations of the general quadratic algebra were studied by Essler and Rittenberg \cite{ER96}, who found explicit representations in terms of infinite dimensional tridiagonal matrices. Quadratic algebras involved in the MPA for coninuous-time processes were studied also in \cite{IPR01}

The matrix-product ansatz marked a breakthrough in the solution of TASEP/ASEP in discrete time under periodic as well as open boundary conditions. For the definition of the different types of discrete-time updates we refer to \cite{RSSS98}. First, by using the MPA, the case of sublattice-parallel update with deterministic bulk dynamics was solved \cite{H96}. The general case of ASEP with stochastic sublattice-parallel dynamics was studied in \cite{HP97}. Next, the TASEP with ordered-sequential update was solved by mapping the corresponding algebra onto the DEHP algebra \cite{RSS96}; see also \cite{RS97}. The case of parallel update (simultaneous updating of all sites) was solved by using two new versions of the matrix-product ansatz. One of these versions leads to a quartic algebra \cite{ERS99}, in contrast to the previous cases, in which the algebra is quadratic. A different representation as a cubic algebra is obtained from a bond-oriented matrix-product ansatz \cite{dGN99}. In general, the MPA has become a powerful method for studying stationary states of different one-dimensional Markov processes out of equilibrium \cite{BE07}. For example, it was used to solve TASEP with a defect particle \cite{M96}, the multi-species TASEP with uniform \cite{EFM09} and inhomogeneous hopping rates \cite{AM13},  the discrete-time case with inhomogeneous rates in the bulk \cite{CMRV15}, ASEP with internal degrees of freedom \cite{CSR10,C13}.

A powerful technique for studying the dynamics of ASEP and some of its generalizations is the Bethe ansatz. According to it, the amplitudes of the eigenfunctions of the transfer matrix can be expressed as a nonlinear combination of properly defined plane waves. This method was used to exactly solve
the open ASEP \cite{GE05}, a three-parametric family of hopping probabilities, which includes TASEP/ASEP \cite{Pov13}, its discrete time versions with inhomogeneous and non local transition rates \cite{CMRV15}, etc. A modified algebraic Bethe ansatz for the continuous-time TASEP with open boundaries, based on results known for integrable quantum spin chains, is given in \cite{C15}.

It should be noted that the properties of the ASEP depend strongly on the choice of the boundary conditions, similarly to the case of systems with long-range interactions. This leads to inequivalence of the nonequilibrium statistical properties for open and closed systems. For instance, the MPA for continuous-time TASEP on a ring becomes trivial: the algebra has Abelian one-dimensional representation \cite{RSSS98}, while in the case of open boundaries the corresponding matrices are generically infinite-dimensional and non-commuting. The open system exhibits (in the thermodynamic limit) three stationary phases in the plane of particle input-output rates, with continuous or discontinuous transitions between them. Another example is the application the Bethe ansatz for obtaining the full current fluctuations in the periodic TASEP \cite{DL98}, and much later in the periodic ASEP \cite{P10}. We mention also, that by using a new form of the Bethe ansatz the totally asymmetric exclusion process on a ring was solved for the non-stationary probabilities under arbitrary initial conditions and time intervals \cite{P03}. The full relaxation dynamics of the TASEP on a ring was solved also by the algebraic Bethe ansatz method \cite{MSS12}. For a review of some results obtained for different versions of ASEP on a ring by using the Bethe ansatz we refer to \cite{M11}.

The steady state of TASEP with parallel update on the ring has a pair-factorized form and exhibits nearest-neighbor correlations \cite{SSNI95}. A factorized form of the steady state of ASEP on a ring was found also in its generalizations including Langmuir kinetics for the attachment and detachment transitions of particles on the chain and in a reservoir, as well in the presence of memory reservoirs \cite{EN12}. For open boundaries, the matrix-product representation was interpreted as a pair-factorized state as on the ring modulated by a matrix-product state \cite{WS09}.

An important, exactly solvable generalization of the TASEP dynamics on a ring was found and studied by W\"{o}lki in 2005 \cite{W05}. That is the exactly solvable representative $pw^{N-1}$ of the general class of so-called $p_1p_2\cdots p_N$-models, under the discrete-time dynamics of which all clusters of particles are updated independently, $p_1p_2\cdots p_N$ being the probability that a $N$-particle cluster will move as a whole one site ahead. Thus, each particle has its own hopping probability $p_i$, $1\le i \le N$, depending on its position in the cluster. It was shown that the model with $p_1=p$ and $p_2 = p_3 = \cdots =
p_{N-1} =: w$ satisfies the condition for models with discrete-mass transport to have a factorized steady state
\cite{EMZ04}. This model, to be denoted here as gTASEP, is the main object of study in the present paper (in our notation $w = \tilde{p}$). We note that the same model was thoroughly studied in the framework of the Bethe integrability in \cite{DPPP}, as a particular case of the general family formulated and investigated in \cite{Pov13}.

Our main aim here is the construction and application of a finite-dimensional matrix-product representation for the gTASEP on finite chains under periodic boundary conditions. The exact expressions for the main characteristics of the stationary state of the model are obtained for arbitrary fixed numbers of the lattice sites $L$ and the particles $N$ ($N\leq L$) on the ring. The correctness of our MPA representation is proved by a parallel combinatorial derivation. The finite-size two-point correlation function is calculated within the MPA.

The paper is organized in six chapters and three appendices. In Sec.~II we formulate the model, Sec.~III presents the matrix-product algebra and its two-dimensional representation. Within a combinatorial approach, in Sec.~IV we calculate the partition function and the average current of particles in the system. Sec.~V  contains derivation of the partition function, the average particle density, the current, and the pair correlation function by using the matrix-product formalism. A discussion of the approach, comparison with known results in several particular regimes of the model, as well as perspectives for further applications are given in Sec.~VI. Explicit results for three small systems are given in Appendix~A, and some details of the calculations are presented in Appendices~B and C. The main results of the study are published in \cite{ABPRE}.

\section{The Model}

We consider TASEP on a ring of $L$ sites, labeled clockwise by the index $i = 1,2,\dots ,L$,
where site $1$ is the nearest-neighbor of site $L$ in the clockwise direction.
Each site of the lattice can be empty or occupied by just one particle.

The dynamics of the model corresponds to the discrete-time backward-ordered update with
probabilities $p$ and $\tilde{p}$ defined as follows. A particle
can hop to a vacant nearest-neighbor site in the clockwise direction, or stay at its place.
During each moment of time $t$, an update of the configuration of the whole system takes place
in $L$ consecutive steps, passing through successive updates of all the pairs of nearest-neighbor sites
in the counterclockwise order $(L-1,L), \dots, (i,i+1),\dots, (1,2), (L,1)$. The probability of a
hop along the bond $(i,i+1)$ depends on whether a particle has jumped from site $i+1$ to
site $i+2$ in the previous step, when the bond $(i+1,i+2)$ was updated, or not.

(1) In the case when the site $i+1$  has not changed its occupation number, the probabilities are
the standard ones: if site $i+1$ remains empty, then the jump of a particle from
site $i$ to site $i+1$ takes place with probability $p$, and the particle stays immobile with
probability $1-p$; if site $i+1$ remains occupied, no jump takes place and the configuration of
the bond $(i,i+1)$ is conserved.

(2) If in the previous step a particle has jumped from site $i+1$ to site $i+2$, thus leaving $i+1$ empty,
then the jump of a particle from site $i$ to site $i+1$ in the next step takes place with a different
probability $\tilde{p}$, and the particle stays immobile with probability $1-\tilde{p}$.

Note that when $\tilde{p} = p$ one has the standard TASEP with backward-sequential update, and
when $\tilde{p} = 0$ one has the TASEP with parallel update.

\section{The matrix-product algebra}

Under the above generalized dynamics, the left-hand site in each pair of nearest-neighbors $(i,i+1)$, which is to be updated,
can be either empty or occupied. With these two states we associate the matrices $E$ and $D$, respectively, and introduce the column-vector
\beq A = \begin{pmatrix} E \\ D \end{pmatrix} .
\label{A} \eeq
On the other
hand, the right-hand site of such a pair $(i,i+1)$ can be in three states: empty, being empty in the previous step of the
update too, occupied, and empty but as a result of a particle hopping from site $i+1$ to site $i+2$ in the previous update step.
With these three states we associate the matrices $\hat{E}$, $\hat{D}$, and $F$, respectively, and introduce the
column-vector
\begin{equation}
\hat{A} = \begin{pmatrix} \hat{E} \\ \hat{D} \\ F \end{pmatrix}
\label{Ahat}.
\end{equation}
Thus, the state of the bond $(i,i+1)$ to be updated is described by the direct matrix product
\beq  A \otimes \hat{A}.
\label{1} \eeq

We conjecture the same update mechanism for each pair of nearest-neighbor sites as in the case of
backward-sequential update, see \cite{RSSS98},
\begin{equation}
\mathcal{T} \left[A\otimes \hat{A}\right] = \hat{A}\otimes A. \label{T}
\end{equation}
With the definition (\ref{Ahat}) of $\hat{A}$, this equation resembles also the stationarity mechanism
suggested for the parallel update in \cite{RSSS98}.

The operator $\mathcal{T}$ in (\ref{T}) is a $6 \times 6$ matrix, defined by the probabilities
of the possible elementary events
\bea
\mathcal{P}(E\hat{E} \rightarrow \hat{E}E)& =& 1, \nn  \\
\mathcal{P}(E\hat{D} \rightarrow \hat{E}D) &= &1, \nn  \\
\mathcal{P}(EF \rightarrow \hat{E}E)& =& 1, \nn   \\
\mathcal{P}(D\hat{E} \rightarrow FD)& = & p, \nn   \\
\mathcal{P}(D\hat{E} \rightarrow \hat{D}E) &= &1-p, \nn \\
\mathcal{P}(DF \rightarrow FD)& = &\tilde{p},  \nn  \\
\mathcal{P}(DF \rightarrow \hat{D}E) &=& 1-\tilde{p},  \nn \\
\mathcal{P}(D\hat{D} \rightarrow \hat{D}D)& = &1.
\label{events} \eea
Here $\mathcal{P}(\cdot)$ denotes the probability of the event in the brackets.
Hence, we obtain the following quadratic algebra:
\bea
\hat{E}E & =& E\hat{E}+ EF,  \label{A1} \\
\hat{E}D &= & E\hat{D} ,  \label{A2} \\
FD & =& \tilde{p}DF + pD\hat{E},  \label{A3} \\
\hat{D}E & = & (1-p)D\hat{E}+(1-\tilde{p})DF,  \label{A4} \\
\hat{D}D  &= & D\hat{D},  \label{A5} \\
FE& = &0 \label{A6}.
\eea
The corresponding stochastic matrix $\mathcal{T}$ has the explicit form
\beq \mathcal{T} =
\begin{pmatrix} 1 & 0 &  1  &  0  &  0  & 0  \\
                0 & 1 &  0  &  0  & 0 & 0 \\
                0 & 0 &  0  & 1-p & 0 &1-\tilde{p}\\
                0 & 0 &  0  & 0  & 1 & 0 \\
                0 & 0 &  0  &  0 & 0 & 0 \\
                0 & 0 &  0  &  p  &  0 & \tilde{p}
\end{pmatrix}
\label{2} \eeq

To solve the quadratic algebra (\ref{A1}) - (\ref{A6}), we make the Ansatz
\beq   \hat{E}=E-F+c_1, \qquad  \hat{D}=D+c_2
\label{Ans}   \eeq
The constants $c_i$, $i=1,2$, will be determined later.
The Ansatz solves trivially Eq. (\ref{A5}), Eq. (\ref{A1}) is satisfied under the
condition of Eq. (\ref{A6}), Eqs. (\ref{A2}) and (\ref{A3}) become equivalent. Thus, the algebra
reduces to the following three equations:
\beq
FD=c_1D-c_2E , \label{A8}  \eeq
\beq pDE - (p-\tilde{p})DF= c_1(1-p)D  + c_2E, \label{A9}  \eeq
\beq    FE=0    \label{A10} .  \eeq

To simplify, we choose $c_2=0$ and obtain
\beq
FD=c_1D,  \label{A11}  \eeq
\beq pDE - (p-\tilde{p})DF= c_1(1-p)D, \label{A12}  \eeq
\beq    FE=0    \label{A13} .  \eeq

A two-dimensional representation of the quadratic algebra  (\ref{A11}) - (\ref{A13}),
depending on four free parameters $d,e,f$, and $c_1$, is provided by the matrices
\bea D= \hat{D}=
 d\begin{pmatrix} 1 &  0  \\ \frac{pe}{f(1-p)}  &  0\end{pmatrix},
\quad
 E =
 \begin{pmatrix}  e & \frac{(p-\tilde{p})f}{p} \\ 0  &  0 \end{pmatrix},
 \nonumber \\
 F=
\begin{pmatrix} 0 & f  \\ 0  & \frac{pe}{1-p}  \end{pmatrix},
\quad
\hat{E} =
\begin{pmatrix} c_1 + e & \frac{(p-\tilde{p})f}{p} \\ 0  &  c_1 - \frac{pe}{1-p} \end{pmatrix}
\label{16}. \eea

One of the most convenient forms of the above representation is obtained when $d=e=f=1$, and
$c_1 =\frac{p}{1-p}$. Then,
\beq D= \hat{D}=
\begin{pmatrix}1 &  0  \\ \frac{p}{1-p}  & 0\end{pmatrix},
\quad
E =
\begin{pmatrix}  1 \; & \frac{p-\tilde{p}}{p} \\ 0  &  0 \end{pmatrix},
\label{2par}
\quad
 F=
\begin{pmatrix} 0 & 1  \\ 0  & \frac{p}{1-p}  \end{pmatrix}.
\eeq
Hence,
\beq \hat{E} =
\begin{pmatrix} \frac{1}{1-p} & -\frac{\tilde{p}}{p}\\ 0  &  0 \end{pmatrix}, \quad
C= E+D = \begin{pmatrix}  2 \; & \frac{p-\tilde{p}}{p} \\
\frac{p}{1-p} &  0 \end{pmatrix}.
\label{2parC} \eeq
Note that, in view of the Ansatz (\ref{Ans}), we have $\hat{C}:=\hat{E}+\hat{D}+F =C+c_1$,
hence, $\hat{C}C=C\hat{C}$.

The most important properties of the representation (\ref{2par}) are
\beq
D^2=D,\quad E^2=E,\quad \operatorname{Tr}(DE)= \frac{1-\tilde{p}}{1-p} \equiv x,\quad \operatorname{det}(DE)=0
. \label{prop} \eeq
In addition, we observe an interesting property of the two-dimensional representation, namely
\beq
 F=\frac {p}{p-\tilde p}(DE-D) = \frac {p}{p-\tilde p}(E-ED)+\frac {p}{1-p}.
\label{Frep}
\eeq
Inserting the expressions for $F$ in terms of the matrices $E$ and $D$ in the algebra (\ref{A11}-\ref{A13}), we obtain the cubic relations:
\beq
EDE =\frac {(1-\tilde p)}{1-p}E, \quad\quad DED=\frac {(1-\tilde p)}{1-p}D.
\label{}
\eeq
Together with $D^2=D, E^2=E$, these relations  amount to  mapping of the quadratic
algebra (\ref{A11}-\ref{A13}) to the Temperley-Lieb algebra $TL_3(x)$, which is thus the symmetry algebra of the gTASEP on a ring.

The eigenvalues of the degenerate matrix $DE$ are $\lambda_1 =0$ and $\lambda_2 =x$. It can be cast in
diagonal form by means of a similarity transformation $V^{-1}DEV$, where
\beq
V^{-1}DEV = \begin{pmatrix} 0 & 0 \\ 0  &  x \end{pmatrix}, \quad V =x^{-1/2}\begin{pmatrix} \frac{p-\tilde{p}}{p} & 1\\ -1  &
\frac{p}{1-p} \end{pmatrix}, \quad V^{-1} =x^{-1/2}\begin{pmatrix}\frac{p}{1-p} & -1\\ 1  &
\frac{p-\tilde{p}}{p} \end{pmatrix}.
\eeq
Hence, $\operatorname{Tr}(DE)^k = x^k$, $k=1,2,3,\dots$, is the crucial feature used in the following consideration.

Thus, the weight $W({\mathcal C})$ of each stationary configuration ${\mathcal C}$, given
by a string of matrices ${\mathcal C}=~DDDEEDEEEE\cdots DDE$, where $D$ stands for occupied site
and $E$ for an
empty one, {\bf depends only on the number of clusters $k({\mathcal C})$ in that configuration:}
\beq W({\mathcal C}) \propto x^{k({\mathcal C})}. \label{W}\eeq

We note that another choice of the parameters, $d=1,\, e=1-p,\, f=1$, and $c_1 =p$, leads to the
representation:
\beq D= \hat{D}=
\begin{pmatrix}1 &  0  \\ p  & 0\end{pmatrix},
\quad
E =
\begin{pmatrix}  1-p \; & \frac{p-\tilde{p}}{p} \\ 0  &  0 \end{pmatrix},
\label{2parH}
\quad
 F=
\begin{pmatrix} 0 & 1  \\ 0  & p  \end{pmatrix},
\eeq
hence
\beq \hat{E} =
\begin{pmatrix} 1 & -\frac{\tilde{p}}{p}\\ 0  &  0 \end{pmatrix}, \quad
C= E+D = \begin{pmatrix}  2-p \; & \frac{p-\tilde{p}}{p} \\
p &  0 \end{pmatrix},
\label{2parCH} \eeq
which has been recently derived by P. Hrab{\'a}k in \cite{PhD}. Note that this
representation does not have all the nice properties (\ref{prop}), because now
\beq D^2=D,\quad E^2= (1-p)E,\quad \operatorname{Tr}(DE)= 1-\tilde{p}. \eeq
However, the matrix $DE$ is degenerate too, with eigenvalues $\lambda_1 =0$ and
$\lambda_2 =1-\tilde{p}$, and the crucial feature $\operatorname{Tr}(DE)^k = (1-\tilde{p})^k$,
$k=1,2,3,\dots$, holds true.
It seems that in this case the weight of a configuration will depend not only
on the number of clusters, but on the number of pairs of holes separating them.
However, this is not the case. Obviously, in a ring of $L$ sites and $N$
particles, there are $L-N$ holes, each of which is represented by an $E$ matrix. Next,
given the number of clusters $k({\mathcal C})\geq 2$ in a configuration ${\mathcal C}$,
the same number of $E$ matrices
are associated with the boundaries $DE$ between the clusters. Hence, the remaining
$f=L-N-k({\mathcal C})$ `free' matrices $E$, each having another $E$ as a left neighbor, will
contribute the factor of $(1-p)^f$, irrespectively of their distribution between the clusters of
particles. Thus, the weight of such configuration will be proportional to:
\beq W({\mathcal C}) \propto (1-p)^{L-N}x^{k({\mathcal C})}.
\label{W1}\eeq
By canceling out the configuration-independent factors $(1-p)^{L-N}$ in the nominator and
the denominator (the partition function $Z(L,N)$), one arrives at the same distribution of
the configuration probabilities as the one given by (\ref{W}).

\section{Combinatorial solution}

Here we calculate the partition function and the average current within a purely combinatorial approach, involving a detailed analysis of all the possible configurations of a given number of particles on a finite periodic lattice,  taking into account their statistical weights (\ref{W1}), and by application of the multinomial theorem. Where possible, we give transparent combinatorial derivation and present the tedious formal derivations in Appendix B.

\subsection{The partition function}

In the matrix-product representation each stationary configuration
is represented as a string of the matrices $E$ (for an empty site) and
$D$ (for an occupied site), e.g. $C = EEEDDEDEE\dots ED$, with a length
equal to the number of lattice sites $L$. Due to the projective properties
of these matrices, the weight $W(C)$ of a configuration $C$ is proportional to
the number of clusters $k(C)$ in the configuration, and does not depend
on any other features of $C$. Thus,
\begin{equation}
W(C)\propto {\mathrm Tr}(DE)^{k(C)} = x^{k(C)}.
\label{WC}
\end{equation}

A more detailed description of $C$ is given by its cluster composition, which is
represented by a partition ${\bf n}(C)$ of the fixed number of particles $N$:
\begin{equation}
{\bf n}(C)= (n_1(C), n_2(C),\dots, n_N(C)): \sum_{j=1}^{N}j n_j(C) = N.
\label{NCn}
\end{equation}
Here ${\bf n}(C)$ is a $N$-component vector with integer coordinates $n_j \geq 0$
denoting the number of clusters of size $j$, such that $\sum_{j=1}^{N}j n_j(C) = N$
(obviously, $n_N \in \{0,1\}$). The total number of clusters in a given configuration
$C$  equals
\begin{equation}
k(C)=  \sum_{j=1}^{N}n_j(C), \quad 1\leq k(C) \leq \min\{N,L-N\} .
\label{kC}
\end{equation}

To calculate the partition function $Z(L,N)$, we have to determine the number of
configurations of $N$ particles with exactly $k$ clusters on the ring of $L$ labeled sites, so
that $1\leq k \leq \min\{N,L-N\}$. To solve the problem, we start by ordering the elements of the partition ${\bf n}_{N,k}$ into an
ordered set ${\bf s}_{N,k}$, corresponding to the clockwise position on the ring of all the $k$ clusters.
Then, to each element of ${\bf s}_{N,k}$ we put into one-to-one correspondence an
initial configuration, constructed as follows. First, we choose any of the $k$ clusters as the
first one, and place its first (leftmost) particle at site 1 of the ring. Next, we realize that the number of
compositions of the $N$ particles into $k$ clusters is
\beq
\left(\begin{array}{c}  N-1 \\ k-1 \end{array}\right),
\label{parcomp}
\eeq
and the composition of the $L-N$ empty sites into the same number $k$ of clusters is
\beq
\left(\begin{array}{c} L- N-1 \\ k-1 \end{array}\right).
\label{empcomp}
\eeq
Obviously, each of the empty clusters can separate any pair of particle clusters consecutively ordered on the ring after the first one, so the number of configurations with fixed position of the first cluster is given by the product of the numbers (\ref{parcomp}) and (\ref{empcomp}). Finally, restoring the translational invariance of the configurations along the ring, we have to multiply the above number by $L$, because the origin can be taken at any site of the ring, and divide it by $k$, because the origin will $k$ times occur at the leftmost site of one the $k$ clusters. Thus, taking into account that the weight of a configuration is given by Eq. (\ref{WC}), we obtain the partition function of the model ($N\ge 1,\, L-N\ge 1$):
\begin{eqnarray}
Z_{L,N}(p,\tilde{p}) & =& L \sum_{k=1}^{\min\{N,L-N\}} x^{k}\left(\begin{array}{c}  L-N-1 \\ k-1 \end{array}\right)\frac{1}{k}\left(\begin{array}{c}  N-1 \\ k-1 \end{array}\right)\nonumber \\ &=&
\frac{L}{L-N}\sum_{k=1}^{\min\{N,L-N\}} \left(\begin{array}{c}  L-N \\ k \end{array}\right)\left(\begin{array}{c} N-1 \\ k-1 \end{array}\right)x^k,
\label{Wsk}
\end{eqnarray}
where $x= (1-\tilde{p})/(1-p)$. In the particular cases when there are no particles or no holes on the ring one has  to set $Z_{L,0}=Z_{L,L}=1$. An independent, but rather involved combinatorial proof of the above result is given in Appendix~B.

Equation (\ref{Wsk}) represents the partition function $Z_{L,N}(p,\tilde{p})$ as a polynomial in
$(1-\tilde{p})/(1-p) := x$. Sometimes it is convenient to express it as a polynomial in
$$\nu = \frac{\tilde{p}-p}{1-p} = 1-x.$$ To this end we expand $x^k =(1-\nu)^k$ according to the binomial
formula and make use of the identity
\begin{equation}
\sum_{k=1}^{\min\{N,L-N\}}\left(\begin{array}{c} L- N-1 \\ k-1 \end{array}\right)\left(\begin{array}{c}  N-1 \\ k-1 \end{array}\right)\frac{1}{k}\left(\begin{array}{c} k \\ m \end{array}\right)= \frac{1}{m}
\left(\begin{array}{c} L- N-1 \\ m-1 \end{array}\right)\left(\begin{array}{c}  L-1- m \\ N-m \end{array}\right)
\end{equation}
and obtain the result
\begin{eqnarray}
Z_{L,N}(p,\tilde{p})&= &L \sum_{m=0}^{\min\{N,L-N\}} \frac{(-\nu)^m(L-m-1)!}{m!(N-m)!(L-N-m)!}
\nonumber \\ &=& \left(\begin{array}{c} L \\ N \end{array}\right){}_2F_1(-N,-L+N;1-L;\nu),
\label{ZLNfin}
\end{eqnarray}
where ${}_2F_1(a,b;c;x)$ is the Gauss hypergeometric function.

Note that the above expression includes the special cases  $Z_{L,0}= Z_{L,L}=1$ and apparently exhibits the particle-hole symmetry $N \leftrightarrow L-N$ of the partition function. In addition, since
\begin{equation}
Z_{L,N}(p,\tilde{p})=  \left(\begin{array}{c} L \\ N \end{array}\right) + L \sum_{m=1}^{\min\{N,L-N\}} \frac{(-\nu)^m(L-m-1)!}{m!(N-m)!(L-N-m)!},
\label{Znu}
\end{equation}
it is convenient to adopt the convention
\begin{equation}
Z_{0,N}(p,\tilde{p})=  \delta_{N,0}.
\label{Z0N}
\end{equation}
When $\tilde{p}=p$ ($\nu =0$), the partition function becomes independent of the jump probability,
\begin{equation}
Z_{L,N}(p,p)=  \left(\begin{array}{c} L \\ N \end{array}\right).
\label{ZLNpp}
\end{equation}

Comparing Eq. (\ref{ZLNfin}) with the result for the corresponding zero-range process (ZRP) obtained in \cite{DPP15}, we see that
\begin{equation}
Z_{L,N}(p,\tilde{p})^{\rm gTASEP} = \frac{L}{L-N}Z_{L,N}(p,\tilde{p})^{\rm ZRP}, \quad N\ge 1,\, L-N\ge 1.
\end{equation}
The factor $L/(L-N)$ is due to the different number of configurations in TASEP and ZRP.

\subsection{The current of particles}

According to the update rules, a cluster of $n$ particles yields the following average number of jumps per
update (unit time):
\begin{equation}
j^{\rm cl}_n = p\left[(1-\tilde{p})\sum_{k=1}^{n-1}k\tilde{p}^{k-1} +n\tilde{p}^{n-1}\right]
= p\left[\sum_{k=1}^{n-1}k\tilde{p}^{k-1} - \sum_{k=1}^{n-1}k\tilde{p}^{k} +
n\tilde{p}^{n-1}\right]=p\sum_{k=1}^{n}\tilde{p}^{k-1}.
\label{Jcl}
\end{equation}

Having in mind the number of different configurations $\mathcal{N}_{\rm diff}({\bf n}_{N,k})$ calculated above for any given partition ${\bf n}_{N,k}$ of the number of particles $N$ into $k$ cluster, by summing up all the contributions we obtain for the current (the average total number of jumps per lattice site)
\begin{eqnarray}
J_{L,N}(p, \tilde{p}) &=&\frac{p}{Z_{L,N}(p,\tilde{p})} \sum_{k=1}^{\min\{N,L-N\}} x^{k}
\left(\begin{array}{c}  L-N-1 \\ k-1 \end{array}\right)\sum_{{\bf n}_{N,k}} \frac{(k-1)!}{\prod_{j=1}^{N}n_j!}
\sum_{s=1}^{N}n_s \sum_{m=0}^{s-1} \tilde{p}^{m} \nonumber \\
&=&\frac{p}{Z_{L,N}(p,\tilde{p})} \sum_{k=1}^{\min\{N,L-N\}} x^{k}\frac{1}{k}
\left(\begin{array}{c}  L-N-1 \\ k-1 \end{array}\right)\sum_{m=0}^{N-1}\tilde{p}^{m}\sum_{{\bf n}_{N,k}}\frac{k!\sum_{s=m+1}^N n_s}
{\prod_{j=1}^{N}n_j!} .
\label{JLN}
\end{eqnarray}
To evaluate the sum
\begin{equation}
\sum_{{\bf n}_{N,k}}\frac{k!}{\prod_{j=1}^{N}n_j!}\sum_{s=m+1}^N n_s
\label{Jsum}
\end{equation}
we apply the operator $$\sum_{s=m+1}^N z^s \frac{\partial}{\partial z^s}$$
to both sides of the multinomial identity (\ref{multi}). The results reads
\begin{equation}
\sum_{{\bf n}_{k}}\frac{k!}{\prod_{j=1}^{N}n_j!}\left[\sum_{s=m+1}^N n_s \right] z_1^{n_1}z_2^{n_2}\cdots z_N^{n_N}=
k \sum_{s=m+1}^N z_s (z_1+z_2+\cdots +z_N)^{k-1}.
\end{equation}
Next, by setting here $z_j=z^j$ $(j=1,2,\dots,N)$, we obtain
\begin{equation}
\sum_{{\bf n}_{k}}\frac{k!}{\prod_{j=1}^{N}n_j!}\left[\sum_{s=m+1}^N n_s\right] z^{\, \sum_j j\, n_j} =
k\sum_{s=m+1}^N z^s (z+z^2+\cdots +z^N)^{k-1}.
\label{Jiden}
\end{equation}
Therefore, when the number of particles is fixed $\sum_j j\, n_j = N$, the sum on the left-hand side of (\ref{Jiden}) at $z=1$
must equal the coefficient of $z^N$ in the expansion of the right-hand side,
\begin{equation}
k[z^N]\sum_{s=m+1}^N z^s (z+z^2+\cdots +z^N)^{k-1}= k\sum_{s=m+1}^N\left(\begin{array}{c}N-s-1 \\ k-2 \end{array}\right)= k\left(\begin{array}{c}N-m-1 \\ k-1 \end{array}\right).
\end{equation}

Thus, the expression for the current (\ref{JLN}) simplifies to
\begin{eqnarray}
&&J_{L,N}(p, \tilde{p}) =\frac{p}{Z_{L,N}(p,\tilde{p})} \sum_{k=1}^{M} x^{k}
\left(\begin{array}{c}  L-N-1 \\ k-1 \end{array}\right)\sum_{m=0}^{N-k}\tilde{p}^{m}\left(\begin{array}{c}N-m-1 \\ k-1 \end{array}\right)\nonumber \\ &&= \frac{x\,p}{Z_{L,N}(p,\tilde{p})}
\sum_{m=0}^{N-1}\tilde{p}^{m}\sum_{k=0}^{M-1} x^{k}\left(\begin{array}{c}  L-N-1 \\ k \end{array}\right)\left(\begin{array}{c}N-m-1 \\ k \end{array}\right),
\end{eqnarray}
where $M\equiv {\min\{N,L-N\}}\ge 1$.

It is instructive to rewrite the above expression as a polynomial in $\nu = 1 - x$. To this end we write $x^k = x\, (1-\nu)^{k-1}$, and expand the last expression according to the Newtonian binomial. By using the identity
\begin{equation}
\sum_{k=0}^{M-1}\left(\begin{array}{c}  k \\ n \end{array}\right) \left(\begin{array}{c}  L-N-1 \\ k \end{array}\right)\left(\begin{array}{c}N-m-1 \\ k \end{array}\right)= \left(\begin{array}{c} L-N-1 \\ n \end{array}\right) \left(\begin{array}{c}  L-m-n-2 \\ N-m-n-1\end{array}\right),
\end{equation}
we obtain
\begin{eqnarray} &&J_{L,N}(p, \tilde{p})=
\frac{x\,p}{Z_{L,N}(p,\tilde{p})}\sum_{m=0}^{N-1}\tilde{p}^{m}\sum_{n=0}^{M-1} \frac{(-\nu)^n}{n!}
\frac{(L-m-n-2)!}{(N-m-n-1)!(L-N-n-1)!}\nonumber \\ &&=
\frac{x\,p}{Z_{L,N}(p,\tilde{p})}\sum_{m=0}^{N-1}\tilde{p}^{m}\left(\begin{array}{c}  L-m-2 \\ N-m-1 \end{array}\right){}_2F_1(1+m-N,N+1-L;2+m-L;\nu)\nonumber \\ &&=
\frac{x\,p}{Z_{L,N}(p,\tilde{p})}
\left(\begin{array}{c}  L-2 \\ N-1 \end{array}\right) F_1(1-N,N+1-L,1,2-L;\nu,\tilde{p})\nonumber \\ &&=
p\left(\frac{1-\tilde{p}}{1-p}\right)\frac{N(L-N)}{L(L-1)}\frac{F_1(1-N,N+1-L,1,2-L;\nu,\tilde{p})}
{{}_2F_1(-N,-L+N;1-L;\nu)} .
\label{Jfin}
\end{eqnarray}
This expression coincides with the TASEP current derived in \cite{DPP15} under a mapping of the ZRP on TASEP.

\section{Matrix-product derivation}

\subsection{The partition function}

According to the Matrix-product Ansatz, the grand canonical partition function is given by
\begin{equation}
Z_L(p, \tilde{p}) = \operatorname{Tr}(C^L),
\label{ZLC}
\end{equation}
with $C=E+D$. To obtain the partition function in the case of fixed number of particles $N$, we introduce a chemical potential $\mu$ of the particles, and define
\begin{equation}
C(\mu) = E +\mu D = \begin{pmatrix}  1+\mu \; & \frac{p-\tilde{p}}{p} \\
\frac{\mu p}{1-p} &  0 \end{pmatrix}.
\label{Cmu}
\end{equation}
Now the partition function for the generalized TASEP, on a ring of $L$ sites with fixed number of particles $N$, we can write in the form
\begin{equation}
Z_{L,N}(p, \tilde{p}) = [\mu^N]\operatorname{Tr}\{C^L(\mu)\},
\label{ZLNC}
\end{equation}
where the symbol $[\mu^N]$ denotes the coefficient of the $\mu^N$ term in the polynomial in $\mu$.  Since the eigenvalues of the 2$\times$2 matrix
$C(\mu)$ are
\begin{equation}
\lambda_{1,2}(\mu;p, \tilde{p} ) = \frac{1}{2}\left[1+\mu \pm \sqrt{(1+\mu)^2 - 4\mu \nu }\right], \label{mu}
\end{equation}
where
\begin{equation}
x = \frac{1-\tilde{p}}{1-p}, \quad \nu := 1-x =\frac{\tilde{p}-p}{1-p}, \label{xmu}
\end{equation}
after some algebra we obtain
\bea
Z_{L,N}(p, \tilde{p})& =& [\mu^N]\left\{\lambda_{1}^L(\mu; p, \tilde{p}) + \lambda_{2}^L(\mu; p, \tilde{p})\right\} \nonumber \\
&=& [\mu^N] 2^{-(L-1)}\sum_{m=0}^{[L/2]} \left(\begin{array}{c}  L\\ 2m\end{array}\right)(1+\mu)^{L-2m}\left[(1+\mu)^2 - 4\mu \nu \right]^m
\nonumber \\
&=&  [\mu^N] 2^{-(L-1)}\sum_{m=0}^{[L/2]} \left(\begin{array}{c}  L\\ 2m\end{array}\right) \sum_{n=0}^m \left(\begin{array}{c}  m\\ n\end{array}\right) 4^n (-\nu)^n\mu^n (1+\mu)^{L-2n}
\nonumber \\
&=&  [\mu^N]2^{-(L-1)}\sum_{m=0}^{[L/2]} \left(\begin{array}{c}  L\\ 2m\end{array}\right) \sum_{n=0}^m \left(\begin{array}{c}  m\\ n\end{array}\right) 4^n (-\nu)^n
\sum_{k=0}^{L-2n} \left(\begin{array}{c} L-2n\\  k\end{array}\right) \mu^{n+k}
\nonumber \\
&=& 2^{-(L-1)}\sum_{m=0}^{[L/2]} \left(\begin{array}{c}  L\\ 2m\end{array}\right) \sum_{n=0}^m \left(\begin{array}{c}  m\\ n\end{array}\right) 4^n (-\nu)^n
\left(\begin{array}{c} L-2n\\ N-n\end{array}\right).
\eea
Next, we change the order of summation over $m$ and $n$, and make use of the identity
\begin{equation}
\sum_{m=n}^{[L/2]} \left(\begin{array}{c}  L\\ 2m\end{array}\right)\left(\begin{array}{c}  m\\ n\end{array}\right)= 2^{L-2n -1}\frac{L}{L-2n}
 \left(\begin{array}{c}  L-n-1\\ n\end{array}\right),
\label{bin}
\end{equation}
to obtain the result
\bea
Z_{L,N}(\tilde{p},p) &=& \sum_{n=0}^{\min\{N,\, L-N\}} \left(\begin{array}{c}  L-n-1\\ n\end{array}\right)\left(\begin{array}{c} L-2n\\N- n\end{array}\right)\frac{L}{L-2n}(x-1)^n
\nonumber \\ &=&  L\sum_{n=0}^{\min\{N,\, L-N\}}(-1)^n \frac{(L-n-1)!}{(N-n)!(L-N-n)!\, n!}\, \nu^n ,
\eea
which is identical with (\ref{ZLNfin}).

\subsection{The local density}

We show here how our matrix-product expression for the particle density $\rho_{L,N} = N/L$ produces several identities involving sums over sets of reduced partition functions. We start from the definition of the average particle density
of the model:
\beq
\rho_{L,N} =  Z_{L,N}^{-1}[\mu^{N-1}]\mathrm{Tr}\left (DC^{L-1}(\mu)\right),
\label{D}
\eeq
which is, obviously, constant over the ring, equal to $N/L$.
The above trace is readily calculated by using the diagonal form of the matrix $C(\mu)$,
\begin{equation}
U^{-1}(\mu)C(\mu)U(\mu) = \begin{pmatrix}  \lambda_1(\mu) \; & 0 \\ 0 & \lambda_2(\mu) \end{pmatrix},
\label{UC}
\end{equation}
where (for the sake of brevity we omit the argument $\mu$)
\begin{equation}
 U = \left[\frac{\tilde{p}-p}{p(\lambda_1-\lambda_2)}\right]^{1/2} \begin{pmatrix}  1 & 1 \\ \frac{p\lambda_2}{\tilde{p}-p} &
\frac{p\lambda_1}{\tilde{p}-p} \end{pmatrix}, \quad U^{-1} = \left[\frac{\tilde{p}-p}{p(\lambda_1-\lambda_2)}\right]^{1/2} \begin{pmatrix}
\frac{p\lambda_1}{\tilde{p}-p} & -1 \\ -\frac{p\lambda_2}{\tilde{p}-p} & 1 \end{pmatrix}.
\label{CmuD}
\end{equation}
Now, the similarity transformation of $D$ with the matrix $U$ yields
\beq U^{-1}DU = \frac{1}{\lambda_1 -\lambda_2}
\left(\begin{array}{cc} \lambda_1 -\nu  & \lambda_1 -\nu \\
\nu - \lambda_2 &  \nu - \lambda_2 \end{array}\right).
\label{B1}
\eeq
Thus,
\bea
\mathrm{Tr}\left (U^{-1}DUU^{-1}C^{L-1}(\mu)U\right )&=&
\frac {1}{\lambda_1-\lambda_2}
\left[(\lambda_1-\nu)\lambda_1^{L-1}-(\lambda_2-\nu)\lambda_2^{L-1} \right]
\nn \\
&=& \left(\sum_{m=0}^{L-1}\lambda_1^m\lambda_2^{L-1-m}
-\nu\sum_{m=0}^{L-2}\lambda_1^m\lambda_2^{L-2-m} \right).
\label{D2}
\eea

Next, we use the equalities,
\bea
\sum_{k=0}^n\lambda_1^k\lambda_2^{n-k}
=\left\{ \begin{array}{l} \sum_{m=0}^{[n/2]}(\mu \nu)^m\left[\lambda_1^{n-2m}+
\lambda_2^{n-2m}\right],\quad n\,= \mathrm{odd} \\
\sum_{m=0}^{n/2 -1}(\mu \nu)^m\left[\lambda_1^{n-2m}+ \lambda_2^{n-2m}\right] + (\mu \nu)^{n/2},
\quad n\, = \mathrm{even}.\end{array}\right. ,
\label{E0}
\eea
and take into account that in the remainder we will need the expression for
\beq
[\mu^q]\sum_{k=0}^n\lambda_1^k\lambda_2^{n-k} = \sum_{m=0}^{[n/2]}\nu^m Z_{n-2m, q-m},
\label{E}
\eeq
which is independent of the parity of $n$. In deriving Eq. (\ref{E}) we have taken into account that for $n$ even one has
$[\mu^q](\mu \nu)^{n/2} = \mu^{n/2}\delta_{q,n/2}$, which equals the summand $\nu^m Z_{n-2m, q-m}$ at the upper limit
$m=n/2$; see (\ref{Z0N}).

Thus, the result for the particle density $\rho_{L,N} = N/L$ can be cast in the form of the identity
\beq
\frac{N}{L}\,Z_{L,N}(\nu)= Z_{L,N}(\nu)+\sum_{m=0}^{[L/2]}\nu^{m}[Z_{L-1-2m,\,N-m-1}(\nu)- Z_{L-2m,\,N-m}(\nu)].
\label{rogen1}
\eeq
An independent proof of this relationship is given in Appendix~C.

Next, we derive a different expression, which follows from the definition of the average density of empty sites:
\beq
1- \rho_{L,N}(\nu) = Z_{L,N}^{-1}[\mu^N]\mathrm{Tr}\left (E C^{L-1}(\mu)\right),
\label{EE}
\eeq
which is, obviously, constant over the ring, equal to $1-N/L$.

The above trace is readily calculated by using the similarity transform of the matrix $E$ with the matrix $U(\mu)$, see Eq. (\ref{CmuD}),
\beq U^{-1}EU = \frac{1}{\lambda_1 -\lambda_2}
\left(\begin{array}{cc} \lambda_1(1- \lambda_2)  & \lambda_1(1 -\lambda_1) \\
- \lambda_2(1-\lambda_2) &  - \lambda_2(1-\lambda_1) \end{array}\right).
\label{BE1}
\eeq
Then
\bea
[\mu^N]\mathrm{Tr}\left (\mu U^{-1}EUU^{-1}C^{L-1}(\mu)U\right )&=&
[\mu^N]\frac {1}{\lambda_1-\lambda_2}
\left[(\lambda_1^L -\lambda_2^L )- \mu \nu(\lambda_1^{L-1}-\lambda_2^{L-1}) \right]
\nn \\
&=&[\mu^N] \left[\sum_{m=0}^{L-1}\lambda_1^m\lambda_2^{L-1-m}
-\mu\nu\sum_{m=0}^{L-2}\lambda_1^m\lambda_2^{L-2-m} \right].
\label{DE2}
\eea

Due to the equality (\ref{E}), the expression for the hole density $1-N/L$ does not depend on the parity of $L$:
\beq
(1-N/L)Z_{L,N}(\nu)= Z_{L,N}(\nu) +\sum_{m=0}^{[L/2]}\nu^m\left[Z_{L-1-2m,\,N-m}(\nu)- Z_{L-2m,\,N-m}(\nu)\right].
\label{rogen1E}
\eeq

Finally, by taking the difference of (\ref{rogen1})  and (\ref{rogen1E}), we find
\beq
(2N/L-1)Z_{L,N}(\nu)= \sum_{m=0}^{[L/2]}\nu^m\left[Z_{L-1-2m,\,N-1-m}(\nu)- Z_{L-1-2m,\,N-m}(\nu)\right].
\label{rogen2E}
\eeq

Thus, here we have established three identities, (\ref{rogen1}), (\ref{rogen1E}), and (\ref{rogen2E}), involving the particle density $N/L$ for all $L\ge 1$ and $1\le N\le L$.

\subsection{The pair correlation function}

The matrix-product form of the two-point particle-particle correlation function
of the model is given by the expression
\bea
G_{L,N}(\tau_i=1,\tau_j=1)&=&[\mu^N]\frac {1}{Z_{L,N}}
{\mathrm Tr}\left (C^{i-1}(\mu)\mu DC^{j-i-1}\mu DC^{L-j}(\mu)\right)\nonumber \\
&=&[\mu^N]\frac {1}{Z_{L,N}}
{\mathrm Tr}\left (\mu DC^{j-i-1}\mu DC^{L-j+i-1}(\mu)\right)
\label{D1}
\eea

Consider first the nearest-neighbor particle-particle correlations, when $j=i+1$. In this case Eq. (\ref{D1}) reduces to
\bea
G_{L,N}(\tau_i=1,\tau_{i+1}=1)&=&[\mu^N]\frac {1}{Z_{L,N}}
{\mathrm Tr}\left (\mu^2 D^2C^{L-2}(\mu)\right )\nn \\
&=&\frac {1}{Z_{L,N}}[\mu^{N-2}]{\mathrm Tr}\left (DC^{L-2}(\mu)\right)= \frac{N-1}{L-1}\,\frac {Z_{L-1,N-1}}{Z_{L,N}}.
\label{nn11}
\eea
Here we have taken into account the relationship $D^2=D$, and used the definition of the
particle density for a ring of length $L-1$ sites having $N-1$ particles, see Eq. (\ref{D}).

Hence, one readily obtains the nearest-neighbor particle-hole correlation function
\bea
&&G_{L,N}(\tau_i=1,\tau_{i+1}=0)=[\mu^N]\frac {1}{Z_{L,N}}
{\mathrm Tr}\left (\mu D E C^{L-2}(\mu)\right )\nn \\
&&=\frac {1}{Z_{L,N}}[\mu^{N}]{\mathrm Tr}\left(\mu DC^{L-1}(\mu)- \mu D \mu D C^{L-2}(\mu) \right)=
\frac{N}{L} - \frac{N-1}{L-1}\,\frac {Z_{L-1,N-1}}{Z_{L,N}}.
\label{nn10}
\eea

It is interesting to note that quite a different in form representation for that function
follows from the direct evaluation of the trace in the first line of Eq. (\ref{nn10}). To this end we make use of the matrices $U(\mu)$ and $U^{-1}(\mu)$ which diagonalize $C(\mu)$, see (\ref{UC}) and (\ref{CmuD}),
to obtain
\begin{equation}
\operatorname{Tr}\{DEC^n\}=\operatorname{Tr}\{U^{-1}DEU {\rm diag}\{\lambda_1^n, \lambda_2^n\}\},
\end{equation}
where ${\rm diag}\{\lambda_1^n, \lambda_2^n\}$ denotes a diagonal 2$\times$2 matrix with eigenvalues $\lambda_1^n$ and $\lambda_2^n$.
Taking into account that
\begin{equation}
 U^{-1}DEU = \frac{\nu}{\lambda_1-\lambda_2} \begin{pmatrix}  (1-\lambda_2)(\lambda_1/\nu -1)\, &\, (1-\lambda_1)(\lambda_1/\nu -1) \\
 (1-\lambda_2)(1- \lambda_2/\nu)\, & \, (1-\lambda_1)(1- \lambda_2/\nu) \end{pmatrix},
\label{DEU}
\end{equation}
we obtain
\begin{eqnarray}
&&\mathrm{Tr}\{DEC^n\}=\nn \\
&&\frac{\nu}{\lambda_1-\lambda_2}\left[-(1+ \lambda_1\lambda_2/\nu)(\lambda_1^n - \lambda_2^n) + \lambda_1\lambda_2(\lambda_1^{n-1} - \lambda_2^{n-1})+(\lambda_1^{n+1} - \lambda_2^{n+1})/\nu\right]\nonumber \\&&=
\frac{\nu}{\lambda_1-\lambda_2}\left[-(1+ \mu)(\lambda_1^n - \lambda_2^n) + \mu \nu(\lambda_1^{n-1} - \lambda_2^{n-1})+(\lambda_1^{n+1} - \lambda_2^{n+1})/\nu\right]\nonumber \\&&=
\nu\left[-(1+\mu)\lambda_2 \sum_{m=0}^{n-1}\lambda_1^m\lambda_2^{n-2-m} + \mu \nu \sum_{m=0}^{n-2}\lambda_1^m\lambda_2^{n-2-m} +
(1/\nu)\sum_{m=0}^{n}\lambda_1^m\lambda_2^{n-m}\right]\nonumber \\&&=
\nu\left[-\mu\nu \lambda_1^{n-1}\lambda_2^{-1} +
(1/\nu -1)\sum_{m=0}^{n-1}\lambda_1^m\lambda_2^{n-m} +(1/\nu)\lambda_1^n \right] \nonumber \\ &&=
x\sum_{m=0}^n \lambda_1^m \lambda_2^{n-m} =\left\{ \begin{array}{l} x \sum_{m=0}^{[n/2]}(\mu \nu)^m\left[\lambda_1^{n-2m}+
 \lambda_2^{n-2m}\right],\quad n\,= \mathrm{odd} \\
x \sum_{m=0}^{n/2 -1}(\mu \nu)^m\left[\lambda_1^{n-2m}+ \lambda_2^{n-2m}\right] +x(\mu \nu)^{n/2},
\quad n\, = \mathrm{even}.\end{array}\right.  \label{TrDE}
\end{eqnarray}
In the above derivation we have used the equalities $x=1-\nu$, $\lambda_1\lambda_2 =\mu \nu$ and $(1+\mu)\lambda_2 = \mu\nu + \lambda_2^2$. Thus, taking into account equality (\ref{E}), we find
\bea
&&G_{L,N}(\tau_i=1,\tau_{i+1}=0)=[\mu^{N-1}]\frac{1}{Z_{L,N}}
{\mathrm Tr}\left (D E C^{L-2}(\mu)\right )\nn \\
&&= \frac{1-\nu}{Z_{L,N}}\sum_{m=0}^{[(L-2)/2]}\nu^m Z_{L-2-2m, N-1-m}=
\frac{N}{L} - \frac{N-1}{L-1}\,\frac {Z_{L-1,N-1}}{Z_{L,N}}.
\label{nn102}
\eea

Now we turn to the pair correlations in the case of general separation between the sites,
$j-i-1=n\ge 1$ and for brevity of notation denote $L-j+i-1 =L-n-2 = m$.
With the use the similarity transformation (\ref{B1}) for $D$, and the diagonal form of the matrix $C$, we
obtain for the trace in expression (\ref{D1})
\bea
&&[\mu^N]{\mathrm Tr}\left (\mu DC^n(\mu)\mu DC^{m}(\mu)\right )= \nn \\
&&= \frac {[\mu^{N-2}]}{(\lambda_1-\lambda_2)^2}
\left[(\lambda_1-\nu)\lambda_1^n-(\lambda_2-\nu)\lambda_2^n\right]
 \left[(\lambda_1-\nu)\lambda_1^{m}-(\lambda_2-\nu)\lambda_2^{m}\right]
\nn \\&&=
\frac{[\mu^{N-2}]}{(\lambda_1-\lambda_2)^2}
\left[\lambda_1^{n+1}-\lambda_2^{n+1}-\nu(\lambda_1^n-\lambda_2^n)\right]
\left[\lambda_1^{m+1}-\lambda_2^{m+1}-\nu
(\lambda_1^{m}-\lambda_2^{m})\right].
\label{B2}
\eea
Next, having in mind that
\beq
\lambda_1^n-\lambda_2^n=(\lambda_1-\lambda_2)
\sum_{k=0}^{n-1}\lambda_1^k\lambda_2^{n-1-k}
\label{B3}
\eeq
and applying equality (\ref{E}), we find
\bea
&&[\mu^{N-2}]{\mathrm Tr}\left(DC^n(\mu) DC^{m}(\mu)\right)=\nn \\
&&[\mu^{N-2}]\left(\sum_{k=0}^n\lambda_1^k\lambda_2^{n-k}-\nu\sum_{k=0}^{n-1}\lambda_1^k\lambda_2^{n-1-k}\right) \left(\sum_{l=0}^m\lambda_1^l\lambda_2^{m-l} -\nu\sum_{l=0}^{m-1}\lambda_1^l\lambda_2^{m-1-l}\right) \nn \\ &&=\sum_{q=0}^{N-2}[\mu^{q}]\left(\sum_{k=0}^n \lambda_1^k\lambda_2^{n-k}-\nu\sum_{k=0}^{n-1}\lambda_1^k\lambda_2^{n-1-k}\right)
[\mu^{N-2-q}]\left(\sum_{l=0}^m\lambda_1^l\lambda_2^{m-l} -\nu\sum_{l=0}^{m-1}\lambda_1^l\lambda_2^{m-1-l}\right) \nn \\
&&=\sum_{q=0}^{N-2}\left[Z_{n+1,q+1}+ \sum_{k=0}^{[(n+1)/2]} \nu^k [Z_{n-2k,\,q-k} -Z_{n+1-2k,\,q+1-k}]
\right]\nn \\
&&\times \left[Z_{m+1,N-1-q} +\sum_{l=0}^{[(m+1)/2]} \nu^l[ Z_{m-2l,\,N-2-q-l} - Z_{m+1-2l,\,N-1-q-l}]
\right].
\label{B4}
\eea
Here we have taken into account that for $n$ even $[n/2]=[(n+1)/2]=n/2$, and for $n$ odd $Z_{n-2k,\,q-k}$ vanishes at the upper limit $k=(n+1)/2$; similarly, for $m$ odd $Z_{m-2l,\,N-2-q-l}$ vanishes at the corresponding the upper limit $l=(m+1)/2$.

The above expression essentially simplifies by noting that the substitution $L=n+1$, $N=q+1$ in
Eq.~(\ref{rogen1}) yields for the summand in the first square brackets in the right-hand side of Eq.~(\ref{B4})
\beq
Z_{n+1,q+1}+ \sum_{k=0}^{[(n+1)/2]} \nu^k [Z_{n-2k,\,q-k} -Z_{n+1-2k,\,q+1-k}] = \frac{q+1}{n+1}\,Z_{n+1,q+1}.
\label{m1}
\eeq
The expression for the summand in the second square brackets in the right-hand side of Eq. (\ref{B4}) follows from (\ref{m1}) under the replacement $n\rightarrow m=L-2-n$, and $q\rightarrow N-2-q$:
\bea
&&Z_{m+1,N-1-q} +\sum_{l=0}^{[(m+1)/2]} \nu^l[ Z_{m-2l,\,N-2-q-l} - Z_{m+1-2l,\,N-1-q-l}]
\nn \\ &&=\frac{N-1-q}{L-1-n}\, Z_{L-1-n,N-1-q}\qquad (m=L-2-n).
\label{m2}
\eea

Thus, we obtain for the pair correlation between particles at sites $i$ and $j=i+1+r$:
\bea
&&G_{L,N}(\tau_i=1,\tau_{i+1+r}=1;\tilde{p},p):=F_{L,N}^{1,1}(r;\nu)\nn \\ &&=\frac{1}{Z_{L,N}(\tilde{p},p)}\sum_{q=\max\{0,r-L+N\}}^{\min\{r,N-2\}}
\frac{q+1}{r+1}\,Z_{r+1,q+1}(\tilde{p},p)\frac{N-1-q}{L-1-r}\,
Z_{L-1-r,N-1-q}(\tilde{p},p)\nn \\
&&=\frac{1}{Z_{L,N}(\nu)}\sum_{q=\max\{0,r-L+N\}}^{\min\{r,N-2\}}
\left(\begin{array}{c} r \\q\end{array}\right)
\left(\begin{array}{c}L-2-r \\ N-2-q\end{array}\right){}_2F_1(-q-1,q-r;-r;\nu)\nn \\
&&\times {}_2F_1(q+1-N,-L+N-q+r;-L+2+r;\nu).
\label{pc}
\eea

This expression is valid for any $r\ge 0$. Indeed, at $r=0$ it reduces to
\beq
F_{L,N}(0;\nu))=\frac{1}{Z_{L,N}(\tilde{p},p)}
Z_{1,1}(\tilde{p},p)\frac{N-1}{L-1}\,Z_{L-1,N-1}(\tilde{p},p),
\eeq
which, in view of $Z_{1,1}(\tilde{p},p)=1$, coincides with the pair correlation function (\ref{nn11}).

Obviously, the pair correlation function (\ref{pc}) is invariant with respect to exchanging the
place of the two distances between the particles on the ring, $r\leftrightarrow L-2-r$.

Remarkably, in the case of the backward sequential update, when $\tilde{p}=p$ ($\nu = 0$), the pair
correlation function $F_{L,N}^{1,1}(r;0)$ becomes constant, independent of both the distance $r$ and the jump probability $p$. Indeed, in this case, taking into account Eq. (\ref{ZLNpp}), we obtain
\bea
&&F_{L,N}^{1,1}(r;p,p)=\nn \\ &&=\frac{1}{Z_{L,N}(p,p)}\sum_{q=\max\{0,r-L+N\}}^{\min\{r,N-2\}}\frac{q+1}{r+1}\,
\left(\begin{array}{c} r+1 \\q+1 \end{array}\right)
\frac{N-1-q}{L-1-r}\,\left(\begin{array}{c}L-1-r \\ N-1-q\end{array}\right)\nn \\
&&=\left(\begin{array}{c} L \\N\end{array}\right)^{-1}\sum_{q=\max\{0,r-L+N\}}^{\min\{r,N-2\}}
\left(\begin{array}{c}r \\ q\end{array}\right)\left(\begin{array}{c}L-2-r \\ N-2-q\end{array}\right)
\nn \\&& =\left(\begin{array}{c} L \\N\end{array}\right)^{-1}\left(\begin{array}{c}L-2 \\ N-2\end{array}\right) = \frac{N(N-1)}{L(L-1)}.
\label{flnbsu}
\eea

Another exact analytic expression follows in the limit $\tilde{p} \rightarrow 1$, i.e., $x = (1-\tilde{p})/(1-p)
\rightarrow 0$. This case models a deterministic (irreversible) aggregation of one-dimensional driven lattice gas. Now, it is convenient to use representation (\ref{Wsk}) for the partition function, since
\beq
Z_{L,N}(\tilde{p}\rightarrow p) = Lx +O(x^2),\quad \mbox{\rm when} \quad N\not= 0,L, \quad \mbox{\rm and} \quad
Z_{L,L}(\tilde{p},p)= 1, \quad L\ge 1.
\eeq
Hence, assuming $N<L$, a nonzero contribution in the sum over $q$ in expression (\ref{pc}) for the pair correlation function will come from the following terms:

(1) The first partition function in the numerator becomes $Z_{r+1,r+1}=1$, and the other multipliers are nonzero. This takes place for all $0\le r \le N-2$, when $q=r$;

(2) The second partition function $Z_{L-1-r,L-1-r}=1$ and all the prefactors are nonvanishing. That occurs for all $L-N \le r\le L-2$, when $q= r -L+N$.

\begin{figure}[ht]
\includegraphics[width=80mm]{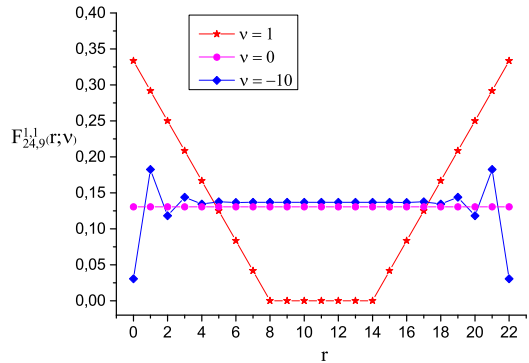} \caption{(Color online) The dependence on the distance $r$ of the particle-particle correlation function  $F_{24,9}^{1,1}(r;\nu)$ on a ring of $L = 24$ sites with $N= 9$ particles: (a) in the limit $\nu \rightarrow 1$, when only configurations containing a single cluster survive (red stars); (b) at $\nu =0$, when the correlations are constant (magenta disks), and (c) at $\nu =-10$, when marked anticorrelations appear at the wings (blue rotated squares).}   \label{F249}
\end{figure}

Thus, we obtain
\beq
F_{L,N}^{1,1}(r;1)=\left\{\begin{array}{ll}(N-1-r)/L,& \quad 0\le r \le N-2\\
(r-L+N+1)/L,& \quad L-N \le r \le L-2 , \end{array}\right.
\label{Fr1p}
\eeq
and $F_{L,N}^{1,1}(r;1)\equiv 0$, if $N-2 <r <L-N$. This behavior is illustrated in Fig. \ref{F249} for the case of $L=24$ and $N=9$. The shape of the correlation function is readily explainable by the fact, that in the limit $\tilde{p} \rightarrow 1$ the configurations of the lowest order in $x \rightarrow 0$ are those in which all the $N$ particles constitute a single cluster.

In the considered limit $\tilde{p} \rightarrow 1$, the value of the nearest-neighbor correlation function is $F_{L,N}^{1,1}(0;1)= (N-1)/L$, and the maximum distance at which non-vanishing particle-particle correlations occur is $r = N-2$, when $F_{L,N}^{1,1}(N-2;1)= 1/L$. The above facts are in exact conformity with the single-cluster stationary state
of the model in the deterministic aggregation regime.

\begin{figure}
\includegraphics[width=80mm]{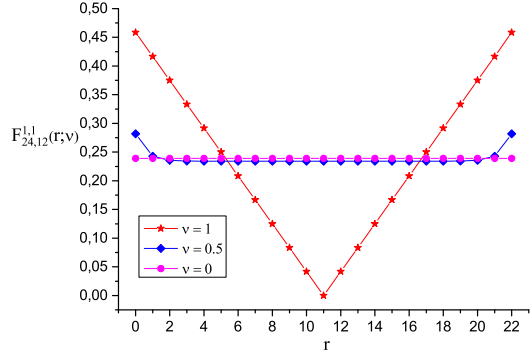} \caption{(Color online) The dependence on the distance $r$ of the particle-particle correlation function  $F_{24,12}^{1,1}(r;\nu)$ on a ring of $L = 24$ sites with $N= 12$ particles: (a) in the limit $\nu \rightarrow 1$, when only single-cluster configurations contribute (red stars), (b) at $\nu = 0.5$, when the graph is considerably more smooth and flat (blue rotated squares), and (c) at $\nu = 0$, when the correlations are constant (magenta disks).}   \label{F2412}
\end{figure}

Another interesting observation concerns the case $2N \ge L+2$, when the summation in Eq. (\ref{pc}) allows
for simultaneous contribution from both partition functions in the numerator. That takes place for distances
$L-N \le r \le N-2$, and the sum of the two results in the right-hand side of Eq. (\ref{Fr1p}) gives the constant value:
Thus, we obtain
\beq
F_{L,N}^{1,1}(r;1)=(2N-L)/L, \quad \mbox{\rm when} \quad L-N \le r \le N-2.
\label{Flat}
\eeq
Actually, this nonzero flat bottom of the pair correlation function extends in the somewhat larger interval
$\quad L-N-1 \le r \le N-1$, since each of the endpoints of that interval contributes the same value,
\beq
F_{L,N}^{1,1}(r=L-N-1;1)=F_{L,N}^{1,1}(r=N-1;1)=(2N-L)/L,
\label{bottom}
\eeq
coming from the second partition function the numerator of expression (\ref{pc}), when $q=r=L-N-1$, and from the first partition function when $r=N-1$ and $q= 2N-L-1$. Thus, the flat bottom at value $(2N-L)/L >0$ occurs
whenever $2N \ge L+1$; when $2N = L$, the function $F_{L,L/2}^{1,1}(r;1)$ is V-shaped, and vanishes at the single-site bottom at $r=L/2 -1$, see Fig. \ref{F2412}.

Obviously, positive values of $\nu < 1$ lead to more smooth and flat graph of the correlation function which,
as $\nu \rightarrow 0$, approaches  the constant value (\ref{flnbsu}) describing a completely uniform distribution of particles in the stationary state, see Fig. \ref{F2412}.

On the other hand, negative values of $\nu$ mark the tendency of splitting clusters of particles into smaller ones by the stochastic dynamics. An extreme case is provided by the parallel dynamics, when $\tilde{p} =0$, hence $\nu = - p/(1-p)$.
Nearest-neighbor particle-hole effective attraction, or particle-particle anticorrelations like those observed on Fig. \ref{F249} at $\nu = -10$, were analytically obtained in the thermodynamic limit for TASEP with parallel update and ring geometry in \cite{SS93}, see also \cite{SSNI95}.

\subsection{The current}

To prove the matrix-product representation for the current (\ref{Jfin}),
\begin{equation}
J_{L,N}(p,\tilde{p})=\frac{p}{Z_{L,N}(p,\tilde{p})}\sum_{k=0}^{N-1}\tilde{p}^k[\mu^{N-k}] \operatorname{Tr}\{\mu DEC^{L-2-k}\},
\label{MPRJ}
\end{equation}
we make use of expression (\ref{TrDE}) for $\operatorname{Tr}\{DEC^n(\mu)\}$ and equality (\ref{E}) for
$[\mu^q]\operatorname{Tr}\{DEC^n(\mu)\}$. For brevity of notation, we set $a=L-2-k$, $b=N-1-k$, and write:
\begin{eqnarray}
&&J_{L,N}(p,\tilde{p})=\frac{p}{Z_{L,N}(p,\tilde{p})}\sum_{k=0}^{N-1}\tilde{p}^k[\mu^{N-k-1}] \operatorname{Tr}\{DEC^{L-2-k}\}
\nonumber \\ &&=\frac{x\,p}{Z_{L,N}(p,\tilde{p})}\sum_{k=0}^{N-1}\tilde{p}^k\sum_{m=0}^{[a/2]}\nu^m [\mu^{b-m}]
\left[\lambda_1^{a-2m}+ \lambda_2^{a-2m}\right]\nonumber \\ &&=\frac{x\,p}{Z_{L,N}(p,\tilde{p})}
\sum_{k=0}^{N-1}\tilde{p}^k\sum_{m=0}^{[a/2]}\nu^m Z_{a-2m,b-m}(p,\tilde{p})
\label{jTr0}
\end{eqnarray}

To bring the above expression to the form of Eq. (\ref{Jfin}), we consider
\bea
&&\sum_{m=0}^{[a/2]}\nu^m Z_{a-2m,b-m}(p,\tilde{p})\nonumber \\ &&=
\sum_{m=0}^{[a/2]}\nu^m (a-2m)\sum_{n\ge 0}\frac{(-\nu)^n(a-2m-n-1)!}{n\,!(b-m-n)!(a-b-m-n)!}\nn \\
&&= \sum_{m=0}^{[a/2]}(-1)^m(a-2m)\sum_{j=0}^{M(a,b)}\frac{(-\nu)^j(a-m-j-1)!}{(j-m)!(b-j)!(a-b-j)!}\nn \\
&&= \sum_{j=0}^{M(a,b)}\frac{(-\nu)^j}{(b-j)!(a-b-j)!}\sum_{m=0}^{j}\frac{(-1)^m (a-2m)(a-m-j-1)!}{(j-m)!}.
\label{jTr}
\eea
Here, in exchanging the order of summation over $m$ and $j$, we have taken into account that
$[a/2]\ge M(a,b)\equiv \min\{a-b,b\}$. Now we prove the equality:
\begin{equation}
\sum_{m=0}^j (-1)^m \frac{(a-2m)(a-1-j-m)!}{(j-m)!}- (-1)^j\delta_{a,2j} = \frac{(a-j)!}{j!}\quad (a \ge 2j),
\label{id1}
\end{equation}
valid for $a \ge 2j$.
The calculation of the sum here is straightforward when we consider the cases of $a \ge 2j+1$ and $a=2j$ separately.

(1) First we assume $a \ge 2j+1$ and set $n= a-2j-1 \ge 0$.  Then the left-hand side of Eq. (\ref{id1}) becomes
\begin{eqnarray}
&&\sum_{m=0}^j (-1)^m \frac{[1+n+2(j-m)](n+j-m)!}{(j-m)!} = \nn \\&&(-1)^j\sum_{m=0}^j (-1)^m \frac{(1+n+2m)(n+m)!}{m!}.
\label{idf0}
\end{eqnarray}
Now we perform the summation as follows:
\begin{eqnarray}
&&(-1)^j\sum_{m=0}^j (-1)^m \frac{(1+n+2m)(n+m)!}{m!}=\nn \\
&&(-1)^j(n+1)!\sum_{m=0}^j (-1)^m \left[\left(\begin{array}{c} n+m\\n\end{array}\right) + 2\left(\begin{array}{c} n+m\\n+1\end{array}\right)\right]
=\nn \\&&(-1)^j(n+1)!\sum_{m=0}^j (-1)^m \left[\left(\begin{array}{c} n+m+1\\n+1\end{array}\right) +
 \left(\begin{array}{c} n+m\\n+1\end{array}\right)\right]=\nn \\
&&(-1)^j(n+1)!\left[-\sum_{m=1}^{j+1}(-1)^m\left(\begin{array}{c} n+m\\n+1\end{array}\right)+\sum_{m=1}^j (-1)^m \left(\begin{array}{c} n+m\\n+1\end{array}\right)\right]=\nn \\
&&\frac{(n+1+j)!}{j!} = \frac{(a-j)!}{j!},
\label{idof}
\end{eqnarray}
which is exactly the right-hand side of Eq.(\ref{id1}).

(2) Finally, let $a=2j$. Then the left-hand side of Eq.(\ref{id1}) becomes
\begin{equation}
\sum_{m=0}^j (-1)^m \frac{2(j-m)(j-1-m)!}{(j-m)!}- (-1)^j = 2\sum_{m=0}^j (-1)^m - (-1)^j =1,
\label{id3}
\end{equation}
which equals the right-hand side of Eq.(\ref{id1}) at $a=2j$. This completes the proof of the equality
under the given condition.

Now, we insert equality (\ref{id1}) for $a>2j$ into Eq. (\ref{jTr}) to obtain:
\beq
\sum_{m=0}^{[a/2]}\nu^m Z_{a-2m,b-m}(p,\tilde{p})=
\sum_{j=0}^{M(a,b)}\frac{(-\nu)^j(a-j)!}{j\,!(b-j)!(a-b-j)!}.
\label{jTr1}
\eeq
Finally, after restoring the original values of $a$ and $b$, and inserting the above result into
Eq.~(\ref{jTr0}), we recover representation (\ref{Jfin}) for the current.

Let us show now, how Eq.~(\ref{jTr0}) reproduces in the thermodynamic limit the well-known result for the
backward-ordered sequential update. When $\tilde{p} =p$, hence $x=1$ and $\nu =0$, Eq.~(\ref{jTr0}) reduces
to
\beq
J_{L,N}(p,p)=\frac{p}{Z_{L,N}(p,p)}\sum_{k=0}^{N-1}p^k Z_{L-2-k,N-1-k}(p,p)
\label{jTrpp}.
\eeq
Note that Eq.~(\ref{ZLNfin}) at $\nu =0$ yields
\beq
Z_{L,N}(p,p)=\frac{L!}{N!(L-N)!} = \left(\begin{array}{c} L\\N \end{array}\right).
\label{Zpp}
\eeq
Therefore, Eq.~(\ref{jTrpp}) can be written as
\bea
&&J_{L,N}(p,p)= p\,\left(\begin{array}{c} L-2\\N-1 \end{array}\right)\left(\begin{array}{c} L\\N \end{array}\right)^{-1}\sum_{k=1}^{N-1} p^k \left(\begin{array}{c} L-2-k\\N-1-k \end{array}\right)\left(\begin{array}{c} L-2\\N-1 \end{array}\right)^{-1}
\nn \\&&=p\,\frac{N(L-N)}{L(L-1)}\left[1+ \sum_{k=1}^{N-1} c_{L,N}(k)p^k \right],
\label{jpp}
\eea
where
\beq
c_{L,N}(k)= \prod_{n=1}^k \frac{N-n}{L-1-n}.
\label{cLNk}
\eeq
Taking the limit $L\rightarrow \infty$ with $N=\rho L$ at fixed $\rho$ and $k$, we obtain
\beq
\lim_{L\rightarrow \infty} c_{L,\rho L}(k)= \rho^k,
\label{ck}
\eeq
and
\beq
\lim_{L\rightarrow \infty}J_{L,\rho L}(p,p)=p\rho(1-\rho)\sum_{k=0}^{\infty} (\rho p)^k =
p\frac{\rho(1-\rho)}{1-p\rho}
\label{jBSU}
\eeq
is the well-known result for the thermodynamic current in the TASEP with backward-ordered update \cite{RSSS98}.

Finally, from the general expression Eq.~(\ref{jTr0}) we derive the well-known result for the
thermodynamic limit of the current in TASEP with parallel update. In this case $\tilde{p} =0$
and Eq.~(\ref{jTr0}) reduces to
\beq
J_{L,N}(p,0)=\frac{p}{Z_{L,N}(p,0)}\sum_{k=1}^{M(L,N)}x^k \left(\begin{array}{c}  L-N-1 \\ k-1 \end{array}\right)\left(\begin{array}{c}  N-1 \\ k-1 \end{array}\right)
\label{jTrp0},
\eeq
where $x =(1-p)^{-1}$, hence $p =1- x^{-1}$. On the other hand, from the expression for the partition
function (\ref{Wsk}) at $\tilde{p} =0$ we have
\beq
Z_{L,N}(p,0)= L\sum_{k=1}^{M(L,N)}\frac{x^k}{k} \left(\begin{array}{c}  L-N-1 \\ k-1 \end{array}\right)\left(\begin{array}{c}  N-1 \\ k-1 \end{array}\right).
\label{Zp0}
\eeq
From Eqs.~(\ref{jTrp0}) and (\ref{Zp0}) it follows that
\beq
J_{L,N}(p,0)=\left. \frac{p\, x}{L}\frac{\partial}{\partial x} \ln Z_{L,N}(1- x^{-1},0)\right|_{x =(1-p)^{-1}}.
\label{Jp0}
\eeq
This remarkable expression allows us to evaluate $\lim_{L\rightarrow \infty}J_{L,\rho L}(p,0)$ by using the
leading order approximation for $Z_{L,\rho L}(p,0)$ as $L \rightarrow \infty$. To this end, we set in
Eq.~(\ref{Zp0}) $N = \rho L$ and $k = yL$, and evaluate the sum by the corresponding Laplace integral:
\beq
Z_{L,\rho L}(p,0)\propto \int_{y=1/L}^{\min\{1-\rho,\rho\}}\mathrm{d}y \exp[L\,S(x,y)],
\label{Zint}
\eeq
where
\bea
S(x,y)&=& (1-\rho)\ln (1-\rho) + \rho \ln \rho + y\ln x - (1-\rho -y)\ln (1-\rho -y)\nn \\
&-& (\rho -y) \ln (\rho -y)- 2y \ln y .
\label{L}
\eea
Hence, one readily finds the equation for the stationary point of $S(x,y)$ as a function of $y$:
\beq
(1-x^{-1})y^2 -y +\rho (1-\rho) =0,
\eeq
and the solution
\beq
y = \bar{y}(x,\rho)\equiv \frac{1-\sqrt{1- 4(1-x^{-1})\rho(1-\rho)}}{2(1-x^{-1})},
\label{ybar}
\eeq
at which $S(x,y)$ attains its maximum with respect to $y$ in the interval $1/L \le y \le \min\{1-\rho,\rho\}$. Thus we obtain
\beq
\lim_{L\rightarrow \infty} \frac{1}{L}\ln Z_{L,\rho L}(p,0)= S(x,\bar{y}(x,\rho)),
\label{limlnZ}
\eeq
which implies
\beq
\lim_{L\rightarrow \infty} J_{L,\rho L}(p,0)=\bar{y}(x =(1-p)^{-1},\rho)= \frac{1-\sqrt{1- 4p\rho(1-\rho)}}{2}.
\label{limJp0}
\eeq
This is exactly the thermodynamic limit for the current of particles in the TASEP with parallel update;
see, e.g., \cite{RSSS98}.

\section{Discussion}

We have studied a version of the TASEP with a generalized discrete-time dynamics described by two hopping probabilities, $p$ and $\tilde{p}$, within the MPA approach to stationary stochastic states. The model is considered on a ring of finite number of sites, labeled in counterclockwise order from 1 to $L$. The configurations with a fixed number of particles $N$ are updated in a cluster-oriented clockwise order, starting with a particle which has a vacant nearest-neighbor site in the hopping direction. If the particle is isolated, or if it is the first particle to be updated in a cluster of particles, then it can jump to its nearest-neighbor site in the counterclockwise direction with probability $p$, and stay immobile with probability $1 - p$. On the other hand, if the particle is not the rightmost one in a cluster of particles (before the update), and if the site in front of it has been emptied in the same update, then the particle can jump ahead with probability $\tilde{p}$, or stay immobile with probability $1-\tilde{p}$. Therefore, $k\ge 1$ particles will chip off a cluster of length $n>k$ with probability $(1-\tilde{p})\tilde{p}^{k-1}p$. Following \cite{DPP15}, we denote the studied model by gTASEP. The gTASEP contains as special cases the TASEP with parallel update, when $\tilde{p} =0$, and with sequential backward-ordered update, when $\tilde{p} =p$. It belongs to the most general class of discrete-mass transport models, and was recognized as its exactly solvable representative $pw^{N-1}$ in the unpublished work \cite{PhD}.

In \cite{PhD} the gTASEP was solved in the thermodynamic limit by using a mean-field theory, and for finite $L$ and $N$ in the partially deterministic case $p=1$. Unfortunately, our matrix representation of the underlying algebra is singular at $p=1$ and we cannot cover that case.

In \cite{DPP15} the stationary properties of the model were obtained by using a mapping onto the zero-range process. The pair correlation function was derived by using the
transfer-matrix method in the grand canonical ensemble with a subsequent choice of the fugacity yielding the prescribed average density of particles in the thermodynamic limit.

In contrast, we have worked entirely in the ensemble with a fixed number of particles on a finite ring, which has lead to more involved calculations. The main aim of the paper was to construct a two-dimensional matrix-product representation for the gTASEP and to use it for the derivation of exact finite-size expressions for the partition function, the current of particles and the two-point correlation function. Our results for the partition function and the current were checked against expressions independently derived by combinatorial methods, as well as by comparison with the results of \cite{DPP15} and the well-known ones for the particular cases of parallel and backward-ordered updates. The obtained expression for the gTASEP partition function is related to the corresponding ZRP by the constant factor $L/(L-N)$, just as the densities and the currents of the two processes are related. This constant factor is due to the different number of configurations in the TASEP and ZRP. Our main new result is the derivation of the finite-size pair correlation function by the MPA method. Its behavior is analyzed in different regimes of effective attraction and repulsion between the particles, depending on whether $\tilde{p} >p$ or $\tilde{p} < p$. In particular, we have explicitly obtained an analytic expression for the pair correlation function in the limit of irreversible aggregation $\tilde{p}\rightarrow 1$, which has confirmed the expectation that in that limit the stationary configurations contain just one cluster.

As a future continuation of the present study we intend to attempt the construction of (infinite-dimensional)
matrix-product representations of the gTASEP on open chains.

\section*{Acknowledgements}

The authors are grateful to A. M. Povolotsky and V. B. Priezzhev for critical reading of the manuscript
and valuable suggestions.

\appendix

\section{Explicit results for small systems}

I. Consider the case of $L=6$, $N=4$. Then the allowed number of clusters is $1\leq k \leq \min\{N,L-N\}=2$.
There is only one partition of 4 into $k=1$ part: ${\bf n}_{4,1}= (0,0,0,1)$, for which
\begin{equation}
\mathcal{N}_{\rm diff}(0,0,0,1)= 6\, S_{0}(1)=6.
\label{N641}
\end{equation}
The partitions of 4 into $k=2$ parts are two: ${\bf n}_{4,2}= (1,0,1,0)$ and ${\bf n}_{4,2}= (0,2,0,0)$.
The first partition yields:
\begin{equation}
\mathcal{N}_{\rm diff}(1,0,1,0)= 6\, S_{1}(0)=6,
\label{N642}
\end{equation}
and the second one:
\begin{equation}
\mathcal{N}_{\rm diff}(0,2,0,0)= 6\, \frac{1}{2!}S_{1}(0)=3.
\label{N643}
\end{equation}
Therefore the partition function is:
\begin{equation}
Z_{6,4}(p,\tilde{p})= 6 \left(\frac{1-\tilde{p}}{1-p}\right) + 9\left(\frac{1-\tilde{p}}{1-p}\right)^2 .
\label{W64}
\end{equation}

For the current we obtain:
\begin{equation}
J_{6,4}(p, \tilde{p}) =\frac{p}{Z(6,4)}\left[(1+\tilde{p}+\tilde{p}^2 + \tilde{p}^3)  \left(\frac{1-\tilde{p}}{1-p}\right) +
(3 + 2\tilde{p}+\tilde{p}^2)\left(\frac{1-\tilde{p}}{1-p}\right)^2\right] .
\label{J64}
\end{equation}

II. Consider the case of $L=7$, $N=4$. Then the allowed number of clusters is $1\leq k \leq \min\{N,L-N\}=3$.
There is only one partition of 4 into $k=1$ part: ${\bf n}_{4,1}= (0,0,0,1)$, for which
\begin{equation}
\mathcal{N}_{\rm diff}(0,0,0,1)= 7\, S_{0}(2)=7.
\label{diff741}
\end{equation}
The partitions of 4 into $k=2$ parts are two: ${\bf n}_{4,2}= (1,0,1,0)$ and ${\bf n}_{4,2}= (0,2,0,0)$.
The first partition yields:
\begin{equation}
\mathcal{N}_{\rm diff}(1,0,1,0)= 7\, S_{1}(1)=14,
\label{N742}
\end{equation}
and the second one:
\begin{equation}
\mathcal{N}_{\rm diff}(0,2,0,0)= 7\, \frac{1}{2!}S_{1}(1)=7.
\label{N743}
\end{equation}
There is only one partition of 4 into $k=3$ parts: ${\bf n}_{4,3}= (2,1,0,0)$, which yields:
\begin{equation}
\mathcal{N}_{\rm diff}(2,1,0,0)= 7\, \frac{2!}{2!}S_{2}(0)=7.
\label{N744}
\end{equation}
Therefore the partition function is:
\begin{equation}
Z_{7,4}(p,\tilde{p}) = 7\left\{ \left(\frac{1-\tilde{p}}{1-p}\right) + 3\left(\frac{1-\tilde{p}}{1-p}\right)^2 +
\left(\frac{1-\tilde{p}}{1-p}\right)^3\right\} .
\label{W74}
\end{equation}

For the current we obtain:
\begin{eqnarray}
J_{7,4}(p, \tilde{p})&=&\frac{p}{Z(7,4)}\left[(1+\tilde{p}+\tilde{p}^2 + \tilde{p}^3)  \left(\frac{1-\tilde{p}}{1-p}\right) +
2 (3 + 2\tilde{p}+\tilde{p}^2)\left(\frac{1-\tilde{p}}{1-p}\right)^2 \right. \nonumber \\
&+&\left.  (3 + \tilde{p})\left(\frac{1-\tilde{p}}{1-p}\right)^3 \right] .
\label{J74}
\end{eqnarray}

III. Consider the case of $L=9$, $N=5$. Then the allowed number of clusters is $1\leq k \leq \min\{N,L-N\}=4$.
There is only one partition of 5 into $k=1$ part: ${\bf n}_{5,1}= (0,0,0,0,1)$, for which
\begin{equation}
\mathcal{N}_{\rm diff}(0,0,0,0,1)= 9\, S_{0}(2)=9.
\label{N951}
\end{equation}
The partitions of 5 into $k=2$ parts are two: ${\bf n}_{5,2}= (1,0,0,1,0)$ and ${\bf n}_{5,2}= (0,1,1,0,0)$.
The first partition yields:
\begin{equation}
\mathcal{N}_{\rm diff}(1,0,0,1,0)= 9\, S_{1}(2)=27,
\label{N952}
\end{equation}
and the second one:
\begin{equation}
\mathcal{N}_{\rm diff}(0,1,1,0,0)= 9\, S_{1}(2)=27.
\label{N953}
\end{equation}
There are also two partitions of 5 into $k=3$ parts: ${\bf n}_{5,3}= (2,0,1,0,0)$ and ${\bf n}_{5,3}= (1,2,0,0,0)$.
The first partition yields:
\begin{equation}
\mathcal{N}_{\rm diff}(2,0,1,0,0)= 9\, \frac{2!}{2!}S_{2}(1)=27,
\label{N954}
\end{equation}
and the second one:
\begin{equation}
\mathcal{N}_{\rm diff}(1,2,0,0,0)= 9\, \frac{2!}{2!}S_{2}(1)=27.
\label{N955}
\end{equation}
Finally, there is only one partition of 5 into $k=4$ parts: ${\bf n}_{5,4}= (3,1,0,0)$, which yields:
\begin{equation}
\mathcal{N}_{\rm diff}(3,1,0,0,0)= 9\, \frac{3!}{3!}S_{3}(0)=9.
\label{N956}
\end{equation}
Therefore the partition function is:
\begin{equation}
Z_{9,5}(p, \tilde{p})= 9\left\{\left(\frac{1-\tilde{p}}{1-p}\right) + 6\left(\frac{1-\tilde{p}}{1-p}\right)^2 +
6\left(\frac{1-\tilde{p}}{1-p}\right)^3 + \left(\frac{1-\tilde{p}}{1-p}\right)^4\right\} .
\label{W95}
\end{equation}

For the current we obtain:
\begin{eqnarray}
J_{9,5}(p, \tilde{p})&=&\frac{p}{Z(9,5)}\left[(1+\tilde{p}+\tilde{p}^2 + \tilde{p}^3 + \tilde{p}^4)  \left(\frac{1-\tilde{p}}{1-p}\right) + 3 (4 + 3\tilde{p}+ 2\tilde{p}^2  + \tilde{p}^3)\left(\frac{1-\tilde{p}}{1-p}\right)^2\right. \nonumber \\ &+& \left.  3 (6 + 3\tilde{p} +\tilde{p}^2)\left(\frac{1-\tilde{p}}{1-p}\right)^3 + (4 +\tilde{p})\left(\frac{1-\tilde{p}}{1-p}\right)^4  \right] .
\label{J95}
\end{eqnarray}

\section{Details of the combinatorial derivations}

Consider first the problem of the combinatorial calculation of the partition function.
We remind the reader that a partition ${\bf n}(C)$ of the fixed number of particles $N$,
\begin{equation}
{\bf n}(C)= (n_1(C), n_2(C),\dots, n_N(C)): \sum_{j=1}^{N}j n_j(C) = N,
\label{NCnB}
\end{equation}
we represent by a $N$-component vector with integer coordinates $n_j \geq 0$
denoting the number of clusters of size $j$.
In the particular case when a configuration contains just one cluster, i.e. $k=1$, rotations
along the ring produce $L$ different configurations with that single cluster. The case
$k\geq 2$ is more involved: if some configuration consists of clusters with different
size, i.e., when for at least two different $l$ and $l'$ one has $n_l\geq 1$ and $n_{l'}\geq 1$,
then different ordering on the ring of the clusters with different size corresponds
to different configurations within the same partition ${\bf n}_{N,k}$ of the number of particles $N$
into $k$ clusters. On the other hand, permutation of clusters with the same size does not change the configuration, since the particles are indistinguishable.

To solve the problem, we `closely pack' all the $k$ clusters, preserving their order on the ring. Thus, we obtain an initial configuration in which cluster $j$ is separated from cluster $j+1$, $j=1,2, \dots k-1$, by exactly one empty site, called the `front bumper' of cluster $j$.
The number of initial configurations, corresponding to the same partition ${\bf n}_{N,k}$
of $N$ into $k$ parts, equals the number to different permutations of $k$ clusters, among which
there are $n_j$, $1\leq j \leq N$, indistinguishable clusters of length $j$:
\begin{equation}
\mathcal{N}_{\rm ini}({\bf n}_{N,k})= \frac{k!}{\prod_{j=1}^{N}n_j!}.
\label{perm}
\end{equation}
Obviously, permutations do not change the length of the close-packed configuration, and the bumper
of the last ($k$-th) cluster always occupies site $N+k$ of the ring. If $N+k = L-q$ with some $q\geq 1$,
then we can generate new configurations with the same composition ${\bf s}_{N,k}$
by keeping the first cluster fixed and translating the $k$-th one as a hole by $m_k=0,\dots, q$ sites
clockwise; then the $(k-1)$-th cluster can be translated by $m_{k-1}=0,1,\dots,m_{k}$ sites, and
so on, until we come up to the second cluster and translate it by $m_{2}=0,1,\dots,m_{3}$ sites. Therefore,
the total number of different configuration with the same composition and fixed position of the first
cluster equals
\begin{eqnarray}
S_{k-1}(q)&=& \sum_{j_{k-1} =0}^{q}\sum_{j_{k-2} =0}^{j_{k-1}}\cdots \sum_{j_1 =0}^{j_{\, 2}}1=
\sum_{n=0}^{k-2} \left(\begin{array}{c}  k-2 \\ n \end{array}\right)\frac{1}{(n+1)!}\prod_{m=0}^{n}(q+1-m)
\nonumber \\&=& \sum_{n=0}^{k-2} \left(\begin{array}{c}  k-2 \\ n \end{array}\right)\left(\begin{array}{c}  q+1 \\n+1 \end{array}\right) = \left(\begin{array}{c}  k+q-1 \\ k-1 \end{array}\right),
\quad k\geq 2, \label{sk}
\end{eqnarray}
with $q=L-N-k$. Some particular values of the above function are:
\begin{equation}
S_{0}(q):= 1,\, S_1(q)=q + 1,\, S_2(q) =\frac{1}{2}(q+1)(q+2),\, S_3(q)=\frac{1}{6}(q+1)(q+2)(q +3).
\label{partsk}
\end{equation}

Next, we consider rotations along the ring of each initial configuration as a whole. Each time, when the first site of cluster $j$, $1\leq j \leq k$ takes position 1 on the ring, an initial configuration appears,
which corresponding to a periodic permutation (including the identical one) of the initial cluster arrangement.
The same holds true for each of the $S_{k-1}(L-N-k)$ configurations, generated from a given initial one by
keeping the first segment fixed and relaxing the position of the remaining $k-1$ clusters. Therefore,
the total number of different configuration, with the same partition ${\bf n}_{N,k}$ of $N$ into $k$ parts,
is generated by the aperiodic permutations of the initial configuration. Taking into account that the $L$ possible positions on the ring of the first segment of each aperiodic configurations produces the factor of $L$, we obtain:
\begin{equation}
\mathcal{N}_{\rm diff}({\bf n}_{N,k})= L\, \frac{(k-1)!}{\prod_{j=1}^{N}n_j!}S_{k-1}(L-N-k)=
L\, \frac{(k-1)!}{\prod_{j=1}^{N}n_j!}\left(\begin{array}{c}  L-N-1 \\ k-1 \end{array}\right) .
\label{diff}
\end{equation}
Now, by using the weight Eq. (\ref{WC}) of each configuration, we obtain for the partition function in the form
\beq
Z_{L,N}(p,\tilde{p})  = L \sum_{k=1}^{\min\{N,L-N\}} x^{k}\left(\begin{array}{c}  L-N-1 \\ k-1 \end{array}\right)\sum_{{\bf n}_{N,k}}\frac{(k-1)!}{\prod_{j=1}^{N}n_j!}.
\label{WskB}
\eeq
where $x= (1-\tilde{p})/(1-p)$ and $\sum_{{\bf n}_{N,k}}$ denotes the sum over all compositions of $N$ into $k$ parts, i.e., $\sum_{j=1}^{N}n_j =k$, and $\sum_{j=1}^{N}j\, n_j=N$, where $n_j \geq 0$ are non-negative integers. To recover the result (\ref{Wsk}) we need to prove the identity
\begin{equation}
\sum_{{\bf n}_{N,k}}\frac{k!}{\prod_{j=1}^{N}n_j!} = \left(\begin{array}{c}  N-1 \\ k-1 \end{array}\right),
\label{iden}
\end{equation}
where the binomial in the right-hand side expresses the number of compositions of $N$ in $k$ parts. That can  be readily done by using multinomial theorem:
\begin{equation}
\sum_{{\bf n}_{k}}\frac{k!}{\prod_{j=1}^{N}n_j!}\,z_1^{n_1}z_2^{n_2}\cdots z_N^{n_N}= (z_1+z_2+\cdots +z_N)^k,
\label{multi}
\end{equation}
where $\sum_{{\bf n}_{k}}$ denotes summation over the non-negative integers $n_j \geq 0$, $j=1,2,\dots N$, under the constraint
$\sum_j n_j =k$. Indeed, by setting here $z_j=z^j$ $(j=1,2,\dots,N)$, we obtain
\begin{equation}
\sum_{{\bf n}_{k}}\frac{k!}{\prod_{j=1}^{N}n_j!}\, z^{\, \sum_j j\, n_j}= (z+z^2+\cdots +z^N)^k.
\end{equation}
Therefore, when $\sum_j j\, n_j = N$, the sum on the left-hand side at $z=1$ must equal the coefficient
of $z^N$ in the expansion of the right-hand side, which is exactly the number of compositions of $N$ in
$k$ parts, found in the right-hand side of identity (\ref{iden}). This completes the proof of the result (\ref{Wsk}).

\section{Proof of the expressions for the particle density}

Expression (\ref{rogen1}) for the particle density has the form
\beq
(1-N/L)Z_{L,N}(\nu)= S_{L,N}(\nu) - \hat{S}_{L-1,N-1}(\nu),
\label{rogen0}
\eeq
where
\bea
&&S_{L,N}(\nu):=\sum_{m=0}^{[L/2]}\nu^m Z_{L-2m,\,N-m}(\nu) \nn \\
&&=\sum_{m=0}^{[L/2]}\nu^m(L-2m) \sum_{n=0}^{M(L,N)} \frac{(-\nu)^n(L-1-2m-n)!}{n!(N-m-n)!(L-N-m-n)!} \nn \\
&&=\sum_{m=0}^{[L/2]}(-1)^m(L-2m) \sum_{p=m}^{M(L,N)}\frac{(-\nu)^p(L-1-m-p)!}{(p-m)!(N-p)!(L-N-p)!},
\label{SLN01}
\eea
and
\bea
&&\hat{S}_{L-1,N-1}(\nu):=\sum_{m=0}^{[L/2]}\nu^m Z_{L-1-2m,\,N-1-m}(\nu) \nn \\
&&=\sum_{m=0}^{[L/2]}\nu^m(L-1-2m) \sum_{n=0}^{M(L-1,N-1)} \frac{(-\nu)^n(L-2-2m-n)!}{n!(N-1-m-n)!(L-N-m-n)!} \nn \\
&&=\sum_{m=0}^{[L/2]}(-1)^m(L-1-2m) \sum_{p=m}^{M(L-1,N-1)}\frac{(-\nu)^p(L-2-m-p)!}{(p-m)!(N-1-p)!(L-N-p)!}.
 \label{SLN02}
\eea
Here $M(L,N)=\min\{L-N,N\}$ and $M(L-1,N-1)=\min\{L-N,N-1\}$. Note that the summand in $\hat{S}_{L-1,N-1}(\nu)$ equals
the one in $S_{L,N}(\nu)$ after the replacement $L\rightarrow L-1$ and $N\rightarrow N-1$ $(L\geq N \geq 1)$. However,
the upper limits in the corresponding sums over $m$ are related in this way only when $L$ is even. In the case of $L$ odd
one has $[(L-1)/2] = [L/2]= (L-1)/2$, and the upper summation limits are the same.

To prove expression (\ref{rogen1}), we first exchange the order of summation over $m$ and $p$ in the above sums. By
noting that $[L/2] \ge M(L,N)\ge M(L-1,N-1)$, we follow the rule
\beq
\sum_{m=0}^{[L/2]}\sum_{p=m}^{M(L,N)}\cdots =\sum_{p=0}^{M(L,N)}\sum_{m=0}^{p}\cdots ,
\label{ge}
\eeq
and obtain:
\bea
S_{L,N}(\nu)
=\sum_{p=0}^{M(L,N)}\frac{(-\nu)^p}{(N-p)!(L-N-p)!}\nn \\ \times
\sum_{m=0}^{p}\frac{(-1)^m (L-2m)(L-1-m-p)!}{(p-m)!}
\label{SLN1}
\eea
and
\bea
\hat{S}_{L-1,N-1}(\nu)
=\sum_{p=0}^{M(L-1,N-1)}\frac{(-\nu)^p}{(N-1-p)!(L-N-p)!} \nn \\ \times
\sum_{m=0}^{p}\frac{(-1)^m (L-1-2m)(L-2-m-p)!}{(p-m)!}.
 \label{SLN2}
\eea

In (\ref{SLN1}) $p\le M(L,N)\le [L/2]$, hence $L\ge 2p$ and the conditions for validity of equality (\ref{id1}) are fulfilled
with $a=L$ and $j=p$. Thus we evaluate the sum
\begin{equation}
\sum_{m=0}^{p}(-1)^m\frac{(L-2m)(L-1-m-p)!}{(p-m)!}= \frac{(L-p)!}{p!}.
\label{idLp}
\end{equation}
The substitution of this result in the right-hand side of expression (\ref{SLN1}) yields
\begin{equation}
S_{L,N}(\nu)= \sum_{p=0}^{M(L,N)}\frac{(-\nu)^{p}}{p\,!}\frac{(L-p)!}{(N-p)!(L-N-p)!}.
\label{SLN11}
\end{equation}

Similarly,
\begin{equation}
\hat{S}_{L-1,N-1}(\nu)= \sum_{p=0}^{M(L-1,N-1)}\frac{(-\nu)^{p}}{p\,!}\frac{(L-1-p)!}{(N-1-p)!(L-N-p)!}.
\label{SLN22}
\end{equation}

Combining the above expressions, we obtain
\bea
&&S_{L,N}(\nu)- \hat{S}_{L-1,N-1}(\nu)= (L-N)\sum_{p=0}^{M(L,N)}\frac{(-\nu)^{p}}{p\,!}\frac{(L-p)!}{(N-p)!(L-N-p)!}
\nn \\&&= (1-L/N)Z_{L,N}(\nu),
\label{PDID}
\eea
which proves equality (\ref{rogen1}).

\end{document}